\documentclass{LMCS}

\def\doi{8(3:27)2012}
\lmcsheading%
{\doi}
{1--25}
{}
{}
{Jul.~25, 2011}
{Sep.~29, 2012}
{}
 
\usepackage{amsfonts}
\usepackage{amsmath}
\usepackage{amssymb}
\usepackage{amstext}
\usepackage{enumerate}
\usepackage{wrapfig}
\usepackage{xspace}
\usepackage{array}
\usepackage{longtable}
\usepackage{multicol}

\usepackage{color}

\usepackage{enumerate,hyperref}

\newcommand\FP{{\rm FP}}
\newcommand\NP{{\rm NP}}

\newcommand\co{{{\rm co}}}

\renewcommand\L{{\rm L}}

\newcommand\leqlogm{\leq^{\rm log}_m}

\newcommand\N{{\mathbb{N}}}	
\newcommand\sset[1]{\{\,#1\,\}}
\newcommand\eqdef{=} 
\newcommand\defeq{}

\renewcommand\qedsymbol{$\Box$}

\def\proof{\trivlist
    \penum=1
    \item[\hskip \labelsep{\it Proof.}]}

\newcount\penum
\def\pitem{\ifnum\penum>1\par\fi\the\penum. \advance\penum by1}

\newlength\problemlength
\newcommand\problemdef[3]{%
\begin{list}{}{\labelwidth\problemlength \labelsep.7em \rightmargin1.5em
\leftmargin\problemlength \advance\leftmargin by3em
\parsep0ex \itemsep.2ex plus.1ex}
\item[{\sl Problem:\hfill}] #1
\item[{\sl Input:  \hfill}] #2
\item[{\sl Output: \hfill}] #3
\end{list}
}

\newlength\dproblength
\newcommand\dproblem[3]{%
\begin{list}{}{\labelwidth\dproblength \labelsep.5em \rightmargin0em
\leftmargin\dproblength \advance\leftmargin by2.3em
\parsep0ex \itemsep.2ex plus.1ex}
\item[{\textsl{Problem:}  \hfill}] #1
\item[{\textsl{Instance:} \hfill}] #2
\item[{\textsl{Question:} \hfill}] #3
\end{list}
}

\newcommand{\var}[1]{\ensuremath{\text{\upshape{Var}}}(#1)}
\newcommand{\sat}{\ensuremath{\mathrm{Sat}}}
\newcommand{\enumsat}[1]{\ensuremath{\mathrm{Enum\ SAT_C}(#1)}}
\newcommand{\FV}[1]{\ensuremath{\mathrm{FV_C}(#1)}}
\newcommand{\USAT}[1]{\ensuremath{\mathrm{Unique\ SAT_C}(#1)}}
\newcommand{\AUDIT}[1]{\ensuremath{\mathrm{AUDIT_C}(#1)}}
\newcommand{\SATstar}[1]{\ensuremath{\mathrm{SAT_C^*}(#1)}}

\newcommand{\Csat}{\ensuremath{\mathrm{SAT_C}}}

\newcommand{\complexityclassname}[1]{\ensuremath{\mathrm{#1}}}
\newcommand{\DP}{\ensuremath{\mathrm{D}}^\ensuremath{\mathrm{P}}}
\newcommand{\coNP}{\complexityclassname{coNP}}
\renewcommand{\P}{\complexityclassname{P}}
\newcommand{\US}{\complexityclassname{US}}

\newcommand{\pc}[1]{\ensuremath{\left[ #1 \right]}}
\newcommand{\enu}[3]{\ensuremath{{#1_{#2}},\allowbreak \dots,\allowbreak#1_{#3}}}

\newcommand{\BF}{{\mathrm{BF}}\xspace}
\newcommand{\Mon}{\ensuremath{\mathrm{M}}\xspace}
\newcommand{\Nmon}{\ensuremath{\mathrm{M_1}}\xspace}
\newcommand{\Omon}{\ensuremath{\mathrm{M_0}}\xspace}
\newcommand{\Tmon}{\ensuremath{\mathrm{M_2}}\xspace}
\newcommand{\Self}{\ensuremath{\mathrm{D}}\xspace}
\newcommand{\Nself}{\ensuremath{\mathrm{D_1}}\xspace}
\newcommand{\Tself}{\ensuremath{\mathrm{D_2}}\xspace}
\newcommand{\Nrep}{\ensuremath{\mathrm{R_1}}\xspace}
\newcommand{\Orep}{\ensuremath{\mathrm{R_0}}\xspace}
\newcommand{\Rep}{\ensuremath{\mathrm{R_2}}\xspace}
\newcommand{\VelC}{\ensuremath{\mathrm{V}}\xspace}
\newcommand{\Tvel}{\ensuremath{\mathrm{V_2}}\xspace}
\newcommand{\Ovel}{\ensuremath{\mathrm{V_0}}\xspace}
\newcommand{\Nvel}{\ensuremath{\mathrm{V_1}}\xspace}
\newcommand{\EtC}{\ensuremath{\mathrm{E}}\xspace}
\newcommand{\Oet}{\ensuremath{\mathrm{E_0}}\xspace}
\newcommand{\Net}{\ensuremath{\mathrm{E_1}}\xspace}
\newcommand{\Tet}{\ensuremath{\mathrm{E_2}}\xspace}
\newcommand{\Sep}{{\mathrm{S}}\xspace}

\newcommand{\ORsep}{\ensuremath{{\mathrm{S}_{02}}}\xspace}

\newcommand{\ORMsep}{\ensuremath{{\mathrm{S}_{00}}}\xspace}
\newcommand{\Nsep}{\ensuremath{{\mathrm{S}_1}}\xspace}
\newcommand{\NRsep}{\ensuremath{{\mathrm{S}_{12}}}\xspace}

\newcommand{\NRMsep}{\ensuremath{{\mathrm{S}_{10}}}\xspace}
\newcommand{\Omsep}[1]{\ensuremath{{\mathrm{S}_0^{#1}}}\xspace}

\newcommand{\Lin}{\ensuremath{{\mathrm{L}}}\xspace}
\newcommand{\Olin}{\ensuremath{{\mathrm{L_0}}}\xspace}
\newcommand{\Nlin}{\ensuremath{{\mathrm{L_1}}}\xspace}
\newcommand{\Tlin}{\ensuremath{{\mathrm{L_2}}}\xspace}
\newcommand{\Rlin}{\ensuremath{{\mathrm{L_3}}}\xspace}
\newcommand{\Neg}{\ensuremath{{\mathrm{N}}}\xspace}
\newcommand{\Tneg}{\ensuremath{{\mathrm{N_2}}}\xspace}
\newcommand{\Ids}{\ensuremath{{\mathrm{I}}}\xspace}
\newcommand{\TIds}{\ensuremath{{\mathrm{I_2}}}\xspace}
\newcommand{\OIds}{\ensuremath{{\mathrm{I_0}}}\xspace}
\newcommand{\NIds}{\ensuremath{{\mathrm{I_1}}}\xspace}

\newcommand{\dua}{{\mathrm{dual}}}
\newcommand{\redpeq}{\ensuremath{\equiv_m^p}\xspace}
\newcommand{\redl}{\ensuremath{\le_m^\mathrm{log}}\xspace}
\newcommand{\redleq}{\ensuremath{\equiv_{\mathrm{m}}^{\mathrm{log}}}}
\newcommand{\loeq}[0]{\ensuremath{\equiv}}
\newcommand{\loiso}[0]{\ensuremath{\cong}}
\newcommand{\PPrefixC}[0]{\ensuremath{\mathrm{C}}}
\newcommand{\parity}{\ensuremath{\oplus}}
\newcommand{\PCEQ}[1]{\ensuremath{\mathrm{EQ_\PPrefixC}(#1)}}
\newcommand{\PCISO}[1]{\ensuremath{\mathrm{ISO_\PPrefixC}(#1)}}

\newcommand{\PCVAL}[1]{\ensuremath{\mathrm{VAL_\PPrefixC}(#1)}}
\newcommand{\TTAUT}{\ensuremath{3\,\text{-}\TAUT}\xspace}
\newcommand{\PCSAT}[1]{\ensuremath{\mathrm{SAT}_\PPrefixC(#1)}}
\newcommand{\ParL}{\ensuremath{\parity\L}\xspace}
\newcommand{\range}[2]{\ensuremath{#1,\allowbreak \dots ,\allowbreak #2}}
 
\newcommand{\gdw}{\ensuremath{\text{ iff }}\xspace}
\newcommand{\boolf}[2]{\ensuremath{f_{#1}(#2)}}
\newcommand{\TDNF}[0]{\ensuremath{3}\,\text{-DNF}\xspace}
\newcommand{\define}{\ensuremath{=}}

\newcommand{\OR}{\ensuremath{\lor}}
\newcommand{\XOR}{\ensuremath{\oplus}}
\newcommand{\NOT}{\ensuremath{\neg}}

\newcommand{\IMP}{\ensuremath{\rightarrow}}
\newcommand{\czero}{0}
\newcommand{\cone}{1}
\newcommand{\ID}{\ensuremath{{\mathit{id}}}}
\newcommand{\TAUT}{\ensuremath{\mathrm{TAUT}}}

\newcommand{\caseDistinction}[1]
           {\left\{ 
            \begin{array}{l@{\quad}l}
              #1
            \end{array} \right. 
           }
\newcommand{\EFV}[1]{\ensuremath{\exists \ \mathrm{FV_C}(#1)}}
\newcommand{\SATP}{\ensuremath{\mathrm{SATP}}}

\newcommand{\dual}[1]{\ensuremath{\mathrm{dual}(#1)}}
\newcommand{\uint}{\ensuremath{\mathbb{N}}}
\newcommand{\set}[1]{\ensuremath\left\{#1\right\}}
\newcommand{\card}[1]{\ensuremath{|#1|}}
\newcommand{\clone}[1]{\ensuremath{\left[ #1\right]}}
\newcommand{\redpm}{\ensuremath{\leq_{m}^{p}}}
\newcommand{\nmodels}{\ensuremath{\not\models}}
\newcommand{\clonename}[1]{\mathrm{#1}}
\newcommand{\suchthat}{\ensuremath{:}}

\newcommand{\cS}{\ensuremath{\clonename{S}}}

\newcommand{\cV}{\ensuremath{\clonename{V}}}

\begin{document}

\title[Complexity classifications for Boolean circuits]{Complexity classifications for different equivalence and audit problems for Boolean circuits}

\author[E.~B\"{o}hler]{Elmar B\"{o}hler\rsuper a}
\address{{\lsuper a}Theoretische Informatik, Universit\"at W\"urzburg, Am Hubland,  D-97030 W\"urzburg, Germany}
\email{boehler@informatik.uni-wuerzburg.de}

\author[N.~Creignou]{Nadia Creignou\rsuper b}
\address{{\lsuper b}Aix-Marseille Universit\'e, CNRS, LIF UMR 7279,  13 288 Marseille, France }
                                                  
\email{creignou@lif.univ-mrs.fr}
\thanks{{\lsuper b}Supported by the Agence Nationale de la Recherche under grant ANR-09-BLAN-0011-01}

\author[M.~Galota]{Matthias Galota\rsuper c}
\address{{\lsuper c}{Elektrobit}, Am Wolfsmantel 46,  D-91058 Erlangen, Germany}
\email{Matthias.Galota@elektrobit.com}

\author[S.~Reith]{Stef\/fen Reith\rsuper d}
\address{{\lsuper d}Theoretische Informatik, FB DCSM, Hochschule RheinMain, Kurt-Schumacher-Ring 18, D-65197 Wiesbaden, Germany}
\email{Steffen.Reith@hs-rm.de}

\author[H.~Schnoor]{Henning Schnoor\rsuper e}
\address{{\lsuper e}Institut f\"ur Informatik, Christian-Albrechts-Universit\"at zu Kiel, Christian-Albrechts-Platz 4, 24118 Kiel}	
\email{schnoor@ti.informatik.uni-kiel.de}

\author[H.~Vollmer]{Heribert Vollmer\rsuper f}
\address{{\lsuper f}Institut f\"ur Theoretische Informatik, Leibniz Universit\"at Hannover, Appelstra{\ss}e 4,  30167 Hannover, Germany}
\email{vollmer@thi.uni-hannover.de}
\thanks{{\lsuper f}Supported by DFG VO 630/6-2}

\keywords{Boolean circuits, complexity classification, isomorphism}
\subjclass{F.2.2}

\begin{abstract}
We study Boolean circuits as a representation of Boolean functions and
consider different equivalence, audit, and enumeration problems. For a
number of restricted sets of gate types (bases) we obtain efficient
algorithms, while for all other gate types we show these problems are
at least NP-hard.
\end{abstract}

\maketitle

\section{Introduction}

The study of Boolean functions is an active research topic since more than one
hundred years. Since the early papers of Shannon \cite{rish42,sha38} and Lupanov
\cite{lup58} in the 1940s and 1950s, Boolean circuits (then called
\emph{switching circuits}) have been used as a computation model for Boolean
functions. The computational complexity theory of Boolean circuits developed
rapidly, see Savage's textbook \cite{sav76}. In the meantime many beautiful
results have been proven, e.\,g., in the area of lower bounds or of algebraic
and logical characterizations of small circuit classes, cf.~\cite{weg87,vol99}.

Another development of equal importance is the search for different
representations (sometimes also called \emph{data
structures}, see, e.g., the books \cite{meinelT98,weg00}) for Boolean functions that may facilitate solving presumably hard
problems. Let us explain this with an example. The well-known satisfiability
problem for propositional logic is known to be NP-complete. This immediately
implies that the problem, given a Boolean circuit $C$, to decide if there is an
input for which $C$ outputs $1$ is NP-complete as well. Thus, using Boolean
circuits as a representation for a Boolean function $f$, to determine if
$f^{-1}(1)$ is not empty appears to be a computationally hard problem. However, if
we represent $f$ by a decision tree, satisfiability can be solved in polynomial
time (in the size of the decision tree). The same holds for ordered binary
decision diagrams and different further types of so called \emph{branching
programs}, see \cite{weg00}. This advantage of course has its price: generally,
Boolean circuits are a much more succinct way of representing Boolean functions.
Nevertheless, since the pioneering work by R.~E.~Bryant, branching programs and
in particular ordered binary decision diagrams have turned out to be a suitable
representation for many application areas such as model checking, VLSI design,
computer-aided design, etc; we refer the interested reader to \cite{weg00} for a
discussion.

In this paper, a different approach is advocated. While it is known that in
general satisfiability for Boolean circuits is NP-complete, there are prominent
easy special cases: For example, if we consider only circuits over a monotone
base, the satisfiability problem admits an efficient solution. Another example
is that of linear circuits (i.\,e., circuits with a base of linear functions).
This phenomenon was studied systematically by H.~R.~Lewis in 1979, who showed that
satisfiability is NP-complete if the base contains or can implement the negation
of implication, i.\,e., the function $x\wedge\neg y$. In all other cases,
satisfiability has a polynomial-time algorithm. This \emph{dichotomy result}
holds for Boolean circuits as well as for propositional formulas. The work of
Lewis has been taken up by Reith and Wagner \cite{rewa05} who examined further
algorithmic problems such as the circuit value problem and the problem of
counting the number of satisfying assignments.

Here we study further important algorithmic tasks for the
representation of Boolean functions by Boolean circuits: First we examine the
\emph{equivalence problem}, i.\,e., the question if two given Boolean circuits
represent logically equivalent Boolean functions, and the \emph{isomorphism
problem}, i.\,e., the problem if two given circuits can be made equivalent
through a permutation of their input variables. 
While these problems are of enormous interest in the area of verification and
model-checking, it should be remarked that also from a theoretical viewpoint
they have a long history: they were studied by Jevons and Clifford in the 19th
century and in particular the isomorphism problem became known as the
``Jevons-Clifford Problem''. 
The isomorphism problem admittedly gains its importance from a more theoretical point
 of view. In complexity theory, isomorphism problems in general are notorious since often
they resist a precise complexity theoretic classification. Most famous of course is graph 
isomorphism, a candidate for an ``intermediate problem'' between P and the NP-complete 
problems. Here we obtain a dichotomy distinguishing the easy from the hard cases for 
isomorphism of circuits, but for the hard problems we only have a hardness result, we 
are not able to prove completeness for a complexity class.

A second group of problems we study concerns so called \emph{frozen variables}.
A variable $x$ is frozen in a Boolean circuit $C$ if $C$ is satisfiable and all
its satisfying assignments give the same Boolean value to $x$. We study the
problem to determine if a given circuit has a variable that is frozen. We also
consider a variant that has become known recently under the name \emph{audit
problem}: this is the problem to decide if a given circuit has a frozen variable
or is unsatisfiable. Originally the audit problem stems from the database area.
One can view the value of a frozen variable as
having been compromised by the results of the query expressed by the circuit. 
This is considered problematic with respect to data security questions (see 
\cite{DBLP:journals/jcss/KleinbergPR03}). The audit problem has further practical 
importance also in VLSI design and testing: here, a frozen variable is a hint for a 
stuck-at fault and hence a manufacturing defect within the circuit.

Finally, we study a variant of the counting problem that is also relevant in practice:
Instead of just determining the number of satisfying assignments we are interested in an
efficient way of producing (\emph{enumerating}) all such assignments. Different
notions of ``efficient'' enumeration have been considered in a paper by Johnson
et al.{} \cite{joyapa88}. We recall these notions here (e.\,g., 
polynomial total time, polynomial delay) and study them in the context of enumerating
solutions of Boolean circuits.

For all these problems we obtain complete complexity
classifications: We determine exactly those circuit bases that make the problems
hard (NP-complete or even harder) and for all remaining bases we present
efficient algorithms solving these problems. 

The organization of the paper is as follows: In the next section we define
Boolean functions and Boolean circuits. We also introduce Post's lattice of all
closed classes of Boolean functions; this lattice will be our main technical
tool to obtain the desired complexity results. In Sect.~\ref{sect:problems} we
formally introduce all algorithmic problems that we will classify. In
Sect.~\ref{sect:eqiso} we then turn to equivalence and isomorphism while in
Sect.~\ref{sect:audit} we study all audit-like problems;
Sect.~\ref{sect:enumeration} contains our results on enumeration. Finally,
Sect.~\ref{sect:conclusion} contains a conclusion and presents some open
problems and future research directions.

\section{Preliminaries}

\subsection{Boolean functions and Post's lattice}\label{subsec:Post} 

A \emph{Boolean function} is an $n$-ary function $f: \{0,1\}^n \rightarrow
\{0,1\}$. In the following we will often use well-known Boolean functions as $0$, $1$, $\land$,
$\lor$, $\neg$, $\oplus$,  $\rightarrow$, the implication function, and the $(k+1)$-ary $k$-threshold function $t_k$ verifying
  $t_k( x_1,\ldots ,
x_{k+1} )=1$ if and only if $\sum_{i=1}^{k+1}x_i\ge k$. 

A \emph{clone} is a set of Boolean functions that is closed under superposition, \emph{i.e.}, it  contains all projections (that is, the
functions $f(a_1, \dots , a_n) = a_k$ for $1 \leq k \leq n$ and $n \in \N$) and is closed under arbitrary composition \cite{poka79,sz86,pip97b,lau06}.
Let $B$ be a
finite set of Boolean functions. We denote by $[B]$ the smallest clone
containing $B$ and call $B$ a \emph{base} for $[B]$.  The set $[B]$ corresponds
to the set of all Boolean functions that can be computed by $B$-circuits (as defined below).
All closed classes of Boolean functions are known,
as is their inclusion structure, which forms a lattice. This lattice
is named after its discoverer E.~Post \cite{pos41}.

The following properties are crucial for the below
definitions of the clones:

\begin{iteMize}{$-$}
   \itemsep 5pt 
    \item $f$ is \emph{$c$-reproducing} if $f(c, \ldots , c) = c$, $c \in\{0,1\}$.
      The  functions $\land$ and $\lor$ are $0$- and $1$-reproducing,
      the binary exclusive or, $\oplus$, is $0$-reproducing, but not
$1$-reproducing, whereas the unary negation ($\neg$)
      is neither $1$- nor $0$-reproducing.
    \item $f$ is \emph{monotonic} if $a_1 \leq b_1, \ldots , a_n \leq b_n$
implies $f(a_1, \ldots , a_n) \leq f(b_1, \ldots , b_n)$. Boolean
      functions built up on composition of only $\land,\lor,0,1$ are monotonic,
like for instance $g(x,y,z) \equiv x \wedge (1 \wedge (y \vee z))$.
    \item $f$ is \emph{$c$-separating of degree $k$} if 
      for all $A \subseteq f^{-1}(c)$ of size $|A|\leq k$ 
      there exists an $i \in \{1, \ldots , n\}$ 
      such that $(a_1, \ldots , a_n) \in A$ implies $a_i = c$, $c \in \{0,1\}$.
      The $(k+1)$-ary  $k$-threshold function $t_k$
      is $1$-separating of degree $k$, but not $1$-separating of degree $k+1$.
 For instance $t_2(x,y,z) \equiv (x \wedge
y) \vee (x \wedge z) \vee (y \wedge z)$, which is the ternary majority function,
is $1$-separating of degree 2.
    \item $f$ is \emph{$c$-separating} if 
      $f$ is $c$-separating of degree $|f^{-1}(c)|$.
      The implication $(x\rightarrow y) \equiv \neg x \vee y$ is $0$-separating.
    \item $f$ is \emph{self-dual} if $f(x_{1},\dots,x_{n}) \equiv \neg f(\neg x_1, \ldots , \neg x_n)$.
      The function $g(x,y,z) \equiv (x \wedge \neg y) \vee (x \wedge \neg z)
\vee (\neg y \wedge \neg z)$ is self-dual.
    \item $f$ is \emph{affine} if $f \equiv x_1 \oplus \cdots \oplus x_n \oplus
c$ with $c \in \{0, 1\}$.
      The function $g(x,y,z) \equiv x \oplus y \oplus z \oplus 1$ is affine and
self-dual.
  \end{iteMize}

	\noindent
  
For a list of all Boolean
clones see Table~\ref{Bases} and for their inclusion structure see
Figure~\ref{platt}. For an extensive introduction to
superposition, Post's Lattice and related problems see
\cite{bcrv03}.
  In the naming of the clones the semantic of single indexes is as follows.
Index 2 indicates that the clone contains no constants at all.
  Index 0 (resp. 1) indicates that the clone contains only the constant 0 (resp.
1) but not 1 (resp. 0).
  Clones with no index contain both constants 0 and 1. The only exceptions to
this convention are the clones $\Self$ and $\Nself$ which
  do not contain any constants at all.
  The index * stands for all valid indexes.
  Clones of particular importance in this paper are: 
  \begin{iteMize}{$-$}
	\itemsep 2pt
  \item the clone of all Boolean functions $\BF = [\land,
\neg]=[\land,\lor,\neg,0,1]$
  \item the monotonic clones $\Mon_*$, e.g.,~$\Mon_2 = [\land, \lor]$, $\Mon
= [\land, \lor,0,1]$
  \item the affine clones $\Lin_*$, e.g.,~$\Lin_2 = [x \oplus y \oplus
z]$, $\Lin = [x \oplus y,0,1]$
  \item the disjunctive clones $\VelC_*$, e.g.,~$\Tvel = [\OR]$, $\VelC =
[\OR,0,1]$
  \item the conjunctive clones $\EtC_*$, e.g.,~$\Tet = [\land]$, $\EtC =
[\land,0,1]$
  \item the $c$-reproducing clones $\Nrep$ (the clone of all $1$-reproducing functions), $\Orep$ ($0$-reproducing functions), $\Rep$ (functions that are both $1$- and $0$-reproducing)
  \item the implication clone $\Sep_0 = [\rightarrow]$
  \item the negated-implication clone $\Sep_1 = [x \wedge \neg y]$
  \item the self-dual clones:  $\Self$ self-dual, $\Nself = \Self \cap \Rep$,
$\Tself = \Self \cap \Mon$
  \item the clones $\Sep_{00} = \Sep_0 \cap \Rep \cap \Mon = [x \vee (y \wedge
z)]$,
  $\Sep_{10} = \Sep_1 \cap \Rep\cap \Mon = [x \wedge (y \vee z)]$,
  $\Sep_{12} = \Sep_1 \cap \Rep = [x \wedge (y \vee \neg z)]$ and
  $\Sep_{02} = \Sep_0 \cap \Rep = [x \vee (y \wedge \neg z)].$
  \item the clones $\Ids_*$ containing only the identity and some constant functions, e.g., $\OIds= [\ID, 0]$
  \end{iteMize}

In the following we will often implicitly refer to the inclusion structure of
Post's lattice. Here are some facts that we will use.

\begin{iteMize}{$-$}
	\itemsep 2pt
  \item The function $x\oplus y \oplus z$ is a function of $\Nself$ since it is
in $\Lin_2$ (see the base given in  Table~\ref{Bases})
 and  $\Lin_2\subset \Nself$.
  \item Similarly the ternary majority function $t_2(x,y,z) \equiv (x \wedge y) \vee (x \wedge z)
\vee (y \wedge z)$ is a function of $\Nself$ since 
it is in $\Tself$ 
 and  $\Tself\subset \Nself$.
\item For all $B$ such that $\Sep_{12}\subset [B]\subseteq \Nrep$ there exists a $k\ge 2$ such
that the threshold function $t_k\in [B]$.
Indeed in this case $[B]$ is either  $\Sep_{12}^k$   for some
$k\ge 2$, or  $\Nrep$ or   
$\Rep$ (which both contain $\Sep_{12}^2$).
  \end{iteMize}

\noindent We will often add some constant $c=0$ or $1$ to a clone $C$
and consider the clone $C' = [C \cup \{c\}]$ generated out of $C$ and $c$. With
Post's lattice one can determine this $C'$ quite easily: It is the
 lowest clone above $C$ that contains $c$, i.e.,  the lowest clone above both
$C$ and $I_c$. 
As a consequence a base of $C'$ is obtained  by a base of $C$ to which we add
the constant $c$. 
 The following list contains identities we will frequently use.

\begin{iteMize}{$-$}
	\itemsep 2pt
  \item   $\BF = [ \Sep_{1}\cup \{1\}]$, thus $\{x\wedge\neg y, 1\}$ is a base
of $\BF$.
  \item  $\Sep_{1} = [ \Sep_{12}\cup \{0\}]$, thus $\{x\wedge (y\vee\neg z),
0\}$ is a base of $\Sep_{1}$.
  \item  $\Nrep = [ \Sep_{12}\cup \{1\}]$, thus $\{x\wedge (y\vee\neg z), 1\}$
is a base of $\Nrep$.
  \item  $\Orep = [ \Sep_{02}\cup \{0\}]$, thus $\{x\vee (y\wedge\neg z), 0\}$
is a base of $\Orep$.
\end{iteMize}

\begin{figure}
\begin{center}
\small
\def\hline{\noalign{\hrule height.1pt}}
\def\Hline{\noalign{\hrule height.8pt}}
\begin{tabular}{lll}
\Hline
\textbf{Class} & \textbf{Definition} & \textbf{Base(s)} \\
\Hline
$\BF$ & all Boolean functions & $\{\land, \NOT\}$  \\ \hline
$\Orep$ & $\{\,f \in \BF\mid f$ is 0-reproducing\,\} & $\{\land, \XOR\}$ \\
\hline
$\Nrep$ & $\{\,f \in \BF\mid f$ is 1-reproducing\,\} & $\{\OR, x \oplus y \oplus
1$\} \\ \hline
$\Rep $ & $\Nrep \cap \Orep$ & $\{\OR, x \wedge (y\oplus z \oplus 1)$\} \\
\hline
$\Mon$ & $\{\,f \in \BF\mid f$ is monotonic\,\} & $\{\land, \OR, \czero, \cone$\}
\\ \hline
$\Nmon$ & $\Mon \cap \Nrep$ & $\{\land, \OR, \cone$\} \\ \hline
$\Omon$ & $\Mon \cap \Orep$ & $\{\land, \OR, \czero$\} \\ \hline
$\Tmon$ & $\Mon \cap \Rep$ & $\{\land, \OR\}$ \\ \hline
$\Sep^n_0$ & $\{\,f \in \BF\mid f$ is 0-separating of degree $n$\,\} & $\{\IMP,
\dua(t_n)\}$ \\ \hline
$\Sep_0$ & $\{\,f \in \BF\mid f$ is 0-separating\,\} & $\{\IMP\}$ \\ \hline
$\Sep^n_1$ & $\{\,f \in \BF\mid f$ is 1-separating of degree $n$\,\} & \{$x
\wedge \overline{y}$, $t_n$\} \\ \hline
$\Sep_1$ & $\{\,f \in \BF\mid f$ is 1-separating\,\} & \{$x \wedge
\overline{y}$\} \\ \hline
$\Sep^n_{02}$ & $\Sep^n_0 \cap \Rep$ & \{$x \vee (y \wedge
\overline{z}), \dua(t_n)\}$ \\ \hline
$\Sep_{02}$ & $\Sep_0 \cap \Rep$ & \{$x \vee (y \wedge \overline{z})$\} \\
\hline
$\Sep^n_{01}$ & $\Sep^n_0 \cap \Mon$ & $\{\dua(t_n), \cone\}$ \\ \hline
$\Sep_{01}$ & $\Sep_0 \cap \Mon$ & $\{x \vee (y \wedge z), \cone\}$ \\ \hline
$\Sep^n_{00}$ & $\Sep^n_0 \cap \Rep \cap \Mon$ & $\{x \vee (y \wedge z),
\dua(t_n)\}$ \\ \hline
$\Sep_{00}$ & $\Sep_0 \cap \Rep \cap \Mon$ & $\{x \vee (y \wedge z)\}$ \\ \hline
$\Sep^n_{12}$ & $\Sep^n_1 \cap \Rep$ & $\{x \wedge (y \vee \overline{z}), t_n\}$
\\ \hline
$\Sep_{12}$ & $\Sep_1 \cap \Rep$ & $\{x \wedge (y \vee \overline{z})\}$ \\
\hline
$\Sep^n_{11}$ & $\Sep^n_1 \cap \Mon$ & $\{t_n, \czero\}$ \\ \hline
$\Sep_{11}$ & $\Sep_1 \cap \Mon$ & $\{x \wedge (y \vee z), \czero\}$ \\ \hline
$\Sep^n_{10}$ & $\Sep^n_1 \cap \Rep \cap \Mon$ & $\{x \wedge (y \vee z), t_n\}$
\\ \hline
$\Sep_{10}$ & $\Sep_1 \cap \Rep \cap \Mon$ & $\{x \wedge (y \vee z)\}$ \\ \hline
$\Self$ & $\{\,f\mid f$ is self-dual\,\} & $\{(x\wedge\overline{y}) \vee (x\wedge\overline{z})
\vee (\overline{y}\wedge\overline{z})\}$ \\ \hline
$\Nself$ & $\Self \cap \Rep$ & $\{(x\wedge y) \vee (x\wedge\overline{z}) \vee (y\wedge\overline{z})\}$ \\
\hline
$\Tself$ & $\Self \cap \Mon$ & $\{(x\wedge y) \vee (y\wedge z) \vee (x\wedge z)\}$ \\ \hline
$\Lin$ & $\{\,f\mid$ $f$ is linear\} &
      $\{\XOR, \cone\}$ \\ \hline
$\Olin$ & $\Lin \cap \Orep$ & $\{\XOR\}$ \\ \hline
$\Nlin$ & $\Lin \cap \Nrep$ & $\{\leftrightarrow\}$ \\ \hline
$\Tlin$ & $\Lin \cap \Rep$ & $\{x \oplus y \oplus z\}$ \\ \hline
$\Rlin$ & $\Lin \cap \Self$ & $\{x \oplus y \oplus z \oplus \cone\}$ \\ \hline
$\VelC$ & $\{\,f\mid$ $f$ is an $\OR$-function or a constant function\} &
      $\{\OR, \czero, \cone\}$ \\ \hline
$\Ovel$ & $[\{\OR\}] \cup [\{\czero\}]$ & $\{\OR, \czero\}$ \\ \hline
$\Nvel$ & $[\{\OR\}] \cup [\{\cone\}]$ & $\{\OR, \cone\}$ \\ \hline
$\Tvel$ & $[\{\OR\}]$ & $\{\OR\}$ \\ \hline
$\EtC$ & $\{\,f\mid$ $f$ is an $\land$-function or a constant function\} &
      $\{\land, \czero, \cone\}$ \\ \hline
$\Oet$ & $[\{\land\}] \cup [\{\czero\}]$ & $\{\land, \czero\}$ \\ \hline
$\Net$ & $[\{\land\}] \cup [\{\cone\}]$ & $\{\land, \cone\}$ \\ \hline
$\Tet$ & $[\{\land\}]$ & $\{\land\}$ \\ \hline
$\Neg$ & $[\{\NOT\}] \cup [\{\czero\}] \cup [\{\cone\}]$ & $\{\NOT, \cone\}$,
$\{\NOT, \czero\}$ \\ \hline
$\Tneg$ &  $[\{\NOT\}]$ & $\{\NOT\}$ \\ \hline
$\Ids$ & $[\{\ID\}] \cup [\{\cone\}] \cup [\{\czero\}]$ & $\{\ID, \czero, \cone\}$
\\ \hline
$\OIds$ & $[\{\ID\}] \cup [\{\czero\}]$ & $\{\ID, \czero\}$ \\ \hline
$\NIds$ & $[\{\ID\}] \cup [\{\cone\}]$ & $\{\ID, \cone\}$ \\ \hline
$\TIds$ & $[\{\ID\}]$ & $\{\ID\}$ \\ \Hline
\end{tabular}
\caption{The list of all Boolean clones with definitions and bases, where $t_n
:= \bigvee^{n+1}_{i=1}\bigwedge^{n+1}_{j=1,j\neq i} x_j$ and
      $\dual{f}(a_1, \dots , a_n) = \neg f(\neg a_1 \dots , \neg a_n)$.}
\label{Bases}
\end{center}
\end{figure}

\begin{figure}
\begin{center}
\includegraphics[scale=0.55]{clones.mps}
\end{center}
\caption{Lattice of all Boolean clones}
\label{platt}
\end{figure}

\subsection{Boolean  circuits}\label{subsec:boolean_circuits}

 Let us now
define the central objects that we deal with in this paper, namely
\emph{Boolean circuits} (see also \cite{vol99}):

\begin{defi}
  Let $B$ be a finite set of Boolean functions. A \emph{Boolean circuit} over
  $B$, or a \emph{$B$-circuit} is a tuple $$C=(V,E,\alpha,\beta,o),$$
  where $(V,E)$ is a finite, acyclic, directed graph, $\alpha\colon
  E\rightarrow\uint$ is an injective function, $\beta\colon V\rightarrow
B\cup\{x_i\ \vert\ i\in\uint\},$ and $o\in V,$ such that the following
conditions hold:

\begin{iteMize}{$-$}
\item If $v\in V$ has in-degree $0,$ then
  $\beta(v)\in\{x_i\ \vert\ i\in\uint\}$, or $\beta(v)$ is a $0$-ary
  function from $B,$

\item if $v\in V$ has in-degree $k>0,$ then $\beta(v)$ is a $k$-ary
  function in $B.$
\end{iteMize}

Nodes in $V$ are also called \emph{gates}. A gate $v$ with
$\beta(v)\in\{x_i\ \vert\ i\in\uint\}$ is called an \emph{input-gate},
and $o$ is called \emph{output-gate}. 
{Later the
function $\alpha$ will be used to specify the order of the
predecessors of a gate.}
\end{defi}

With $\var C$ we denote the variables appearing in the circuit $C$,
i.e., the set $\{\beta(v)\ \vert\ v\in
V\}\cap\{x_i\ \vert\ i\in\uint\}$.

This definition of a Boolean circuit corresponds to the intuitive idea
that a circuit consists of a set of gates which are either input
gates, or compute some Boolean function (in our case, functions from
$B$) with arguments taken from the predecessor gates. The set $B$ is
also called a \emph{base}. The distinguished gate $o$ is the
output-gate, i.e., the value computed by the circuit is the result
computed in this gate. The \emph{size} of a circuit is the number of
non-input gates. The function computed by a circuit is defined in the
canonical way: Once we know the values for the input-gates, we can
inductively (since the graph is acyclic) compute the value for each
gate $g\in V$. For non-commutative functions in $B$, the ordering
$\alpha$ on the edges in the graph gives a well-defined function
value. The following definition captures this formally:

\begin{defi}
Let $C=(V,E,\alpha,\beta,o)$ be a Boolean circuit with $\var
C=\{x_1,\dots,x_n\}$, and let $a_1,\dots,a_n\in\set{0,1}.$ Let $v$ be
a gate in $C$. We define the function $f_v$ computed by the gate $v$
on input $(a_1,\dots,a_n)$ as follows:

\begin{iteMize}{$-$}
\item If $v$ is an input-gate, i.e., $\beta(v)=x_i$ for
  $i\in\set{1,\dots,n},$ we define $f_v(a_1,\dots,a_n) \eqdef a_i.$

\item If $v$ has in-degree $k$, and $v_1,\dots,v_k$ are the
  predecessor gates of $v$ in $C$ such that 
  $\alpha\left((v_1,v)\right)<\dots<\alpha\left((v_k,v)\right),$
  then $$f_v(a_1,\dots,a_n)\eqdef\beta(v)(f_{v_1}(a_1,\dots,a_n),\dots,
  f_{v_k}(a_1,\dots,a_n)).$$
\end{iteMize}

\noindent We define the function $f_C\colon\set{0,1}^n\rightarrow\set{0,1}$, the
\emph{function computed by $C$}, as $f_{o}$.
\end{defi}

If the function $f_C$  associated with a circuit $C$ does not depend on
its $i$th argument, we say that $x_i$ is a \emph{fictive} or
\emph{irrelevant} variable for the circuit.\medskip

By writing $C(x_1,\dots, x_n)$, we mean that $C$ is a circuit such that
$\var C\subseteq\{x_1,\dots, \allowbreak x_n\}$. For constant values $a_1,\dots,a_n \in
\sset{0,1}$, we also denote $f_C(a_1,\dots,a_n)$ by
$C(a_1,\dots, \allowbreak a_n)$. An \emph{assignment} for the variables in
$C(x_1,\dots,x_n)$ is a function
$I\colon\{x_1,\dots,x_n\}\rightarrow\{0,1\}$. Such an assignment
is also called \emph{compatible with $C$}. We will write $I\models C$ if
$C(I(x_1),\dots, \allowbreak I(x_n))=1$. In this case we also say that $I$ is a
\emph{satisfying assignment} or a \emph{solution} for $C$. We denote by
$\sat(C)$ the set of assignments satisfying $C$ and by $\#\sat(C)$ the
cardinality of this set.  A circuit
is \emph{satisfiable} if it has a satisfying assignment. When the
order of variables is clear from the context, we write an assignment
simply as the tuple of binary values, i.e., with $(a_1,\dots,a_n)$ we
denote the corresponding assignment $I$ where $I(x_i)=a_i$ for all
relevant $i$.
 For convenience we also often write $C(a_1\dots a_n)$
when we mean $C(a_1,\dots,a_n)$. Hence $C(1^n)$ denotes the value
$C(1,\dots,1)$.
 We sometimes view the circuit as a function of its
assignments, and write $C(I)=1$ if $I\models C$, and $C(I)=0$
otherwise. In the following, let $n \in \uint$, let $f$ be an $n$-ary
Boolean function, $a \in \sset{0,1}$, and $I_1,I_2$ be functions
$I_1,I_2\colon\{x_1,\dots,x_n\}\rightarrow\{0,1\}$. We define $\#_a(I)
\define \#\sset{i \suchthat 1 \le i \le n \text{ and } I(x_i)=a}$.
For assignments $I_1$ and $I_2$, we write $I_1 \leq I_2$ if $I_1(x_i)\leq
I_2(x_i)$ for $1\leq i \leq n$. Finally let $\dual{I}$ be the assignment
$\dual{I}(x_i)=1-I(x_i)$.

For a circuit $C$ and a variable $x\in\var C$, the variable $x$ is
said to be \emph{frozen} in $C$ if $C$ is satisfiable and there is a
constant $c\in\set{0,1}$ such that for all assignments $I$, $I\models C$
implies $I(x)=c$. Similarly we define that $V\subseteq\var C$ is
\emph{frozen} in $C$ if every variable in $V$ is frozen in $C$.

\section{Problems for propositional circuits and complexity classes}
\label{sect:problems}

We now define the list of computational problems involving Boolean circuits
that we study in this paper. All our problems are connected to the
satisfiability problem and the circuit value problem defined as follows---in the following,
let $B$ be a base, i.e., a finite set of Boolean functions.

\dproblem{\Csat{B}}
{A $B$-circuit $C(\enu{x}{1}{n})$  }
{Is $C$ satisfiable?}

\dproblem{\PCVAL{B}}
{A $B$-circuit $C(\enu{x}{1}{n})$ and an assignment $(\enu{a}{1}{n})$}
{Is $\boolf{C}{\enu{a}{1}{n}} = 1$?}

The complexity of these problems is well known:
\begin{prop}
\label{prop:Csat}(\cite{lew79})
Let $B$ be a finite set of Boolean functions. Then $\PCSAT{B}$ is $\NP$-complete
if
$\Nsep\subseteq\pc B$, and solvable in $\P$ otherwise.
\end{prop}

\begin{prop}[\cite{lad75},\cite{rewa05}]
\label{prop:Circval}
Let $B$ be a finite set of Boolean functions, then $\PCVAL{B} \in \P$. 
\end{prop}

We will be   interested in equivalence and isomorphism problems. Let us
 define precisely these two notions. 
\begin{defi}
\label{def:PermAss}
Let $\pi \colon \sset{\range{x_1}{x_n}} \rightarrow \sset{\range{x_1}{x_n}}$
be a permutation and $I\colon
\sset{\enu{x}{1}{n}} \rightarrow \sset{0,1}$ be a truth assignment.
We define the permuted
assignment $\pi(I)$ by $\pi(I)(x_i)\define I(\pi(x_i))$ for $i=1,\ldots , n$.
\end{defi}

\begin{defi}
 Let $C_1(x_1,\dots,x_n)$ and $C_2(x_1,\dots,x_n)$ be $B$-circuits.

The two circuits are {\rm equivalent}, denoted by $C_1 \loeq C_2$, if for all
truth
assignments $I \colon
 \sset{x_1,\dots,x_n} \rightarrow \sset{0,1}$, $I\models C_1$ if and only if
$I\models C_2$.

The two circuits are  {\rm isomorphic}, denoted by $C_1 \loiso C_2$ if  there
exists a
permutation  $\pi \colon
 \sset{x_1,\dots,x_n} \rightarrow \sset{x_1,\dots,x_n}$ such that for all truth 
assignments $I \colon
 \sset{x_1,\dots,x_n} \rightarrow \sset{0,1}$, $I\models C_1$ if and only if
$\pi(I)\models C_2$.
\end{defi}

Using these equivalence relations, we define the \emph{Boolean equivalence} 
and \emph{Boolean isomorphism} problem for
$B$-circuits:
\medskip

\dproblem{\PCEQ{B}}
{Two $B$-circuits $C_1$ and  $C_2$}
{Is $C_1 \loeq C_2$?}

The equivalence problem for propositional circuits or formulas is one
of the standard \coNP-complete problems. The complexity of the next
problem we consider, the isomorphism problem, is not completely
determined. It is clearly \coNP-hard and lies in the second level of
the polynomial hierarchy, more precisely in $\Sigma^p_2$. However, it
is not known to be solvable in \coNP, and is not complete for
$\Sigma^p_2$, unless the polynomial hierarchy
collapses~\cite{agth00}. We study the version of this problem where
the inputs are restricted to $B$-circuits:

\dproblem{\PCISO{B}}
{Two $B$-circuits $C_1$ and  $C_2$}
{Is $C_1 \loiso C_2$?}
\medskip

The next two problems are concerned with frozen variables. As defined
earlier a variable $x$ is frozen in a satisfiable circuit if all
solutions of the circuit assign $x$ the same Boolean value. The
problem of recognizing frozen variables in Boolean formulas was first
studied by Jon Kleinberg, Christos Papadimitriou, and Prabhakar
Raghavan in \cite{DBLP:journals/jcss/KleinbergPR03}; their motivation to consider this problem
was to ensure that database queries do not reveal information that
should be kept secret. Again, we consider the version of two problems
in this context where we restrict the propositional gates allowed to
appear in the input circuits:

\dproblem{$\FV{B}$} {A $B$-circuit $C$ over a set of variables $V$
  and $V' \subseteq V$ such that $|V'| \geq 1$} {Is $V'$ frozen in
  $C$?}

\noindent
If we restrict the problem $\FV {B}$ to instances with $V'=V$, then we
obtain the generalized \textit{Unique Satisfiability problem} over
circuits. This is a natural complete problem for the class
\US\ (see~\cite{blgu82}). We define the problem $\USAT{B}$ to be the
restriction of this problem to $B$-circuits as input.

\dproblem{$\USAT{B}$} {A $B$-circuit, $C$} {Does $C$ have exactly one
  satisfying assignment? }

\noindent The question of the existence of such a frozen variable is the
following problem
\dproblem{$\EFV{B}$}{A $B$-circuit $C$} {Is there a frozen variable in $C$?}

\noindent Note that in the above problem $\EFV B$  it is necessary that the
circuit is satisfiable. If we drop
the restriction of being satisfiable, we have the definition of the so
called \emph{Audit problem}: {Does $C$ have a
frozen variable or is $C$ unsatisfiable? }

\medskip
 Besides these decision problems we are also interested in the enumeration
problem which asks, for a given Boolean circuit to generate the set
of its satisfying assignments with no repetition.
\problemdef{$\enumsat{B}$}{A $B$-circuit, $C$}{All satisfying
assignments of $C$}

In the following in establishing the   complexity of the decision problems
defined above   we need
notions of the following complexity classes: Let $\P$ ($\NP$ resp.) be
the class of languages which are decidable (acceptable, resp.) by
deterministic (nondeterministic, resp.) Turing machines in polynomial
time. For an arbitrary complexity class $\mathcal K$, let ${\mathrm
co}\mathcal{K} \define \sset{\overline{A} \suchthat A \in \mathcal K}$.
Recall that $\DP \define \sset{L \cap L' \suchthat L\in\NP,
L'\in\coNP}$, which is the second level of the Boolean hierarchy and
contains both $\NP$ and $\coNP$ (see \cite{CGH88,CGH89}).

 For our hardness results we mostly employ \emph{logspace many-one reductions}, defined
as follows:
  A language $A$ is logspace many-one reducible to some language $B$ (written $A
\leqlogm B$) if
  there exists a logspace-computable function $f$ such that $x \in A$ if and
only if  $f(x) \in B$. We write $A\redleq B$ if $A\redl B$ and $B\redl A$.
\emph{Polynomial-time many-one reductions} (written as $A\redpm B$ and $A\redpeq B$)
are defined in the same way, except that the function $f$
is only required to be computable in polynomial time.

\medskip
 For the enumeration problem polynomial time is not
a suitable notion of efficiency, since the number of solutions may
be exponential in the length of the circuit. For the notion of an
 ``efficient'' enumeration algorithm, we use the
definitions from \cite{jpy88}.
An algorithm for the enumeration
problem has \emph{polynomial total time}, if the running time of the algorithm
is
polynomial in the length of the input circuit and in the number of its
satisfying solutions. This notion is also referred to as \emph{output
polynomial}. An important feature of an enumeration algorithm is the ability to
start generating solutions 
as soon
as possible, and more generally to generate solutions in a regular way with
a limited delay between two successive outputs.  
 It has \emph{polynomial delay} if the time
needed by the algorithm between its start and the printing of the
first solution, the time between the printing of each two consecutive
solutions, and the time between printing the last solution and the termination
of the algorithm is bounded by a polynomial in the length
of the input circuit. 
In \cite{jpy88} the authors exhibited polynomial-delay algorithms that used exponential space and therefore distinguished polynomial-delay algorithms using only polynomial space. In our paper  polynomial-delay enumeration algorithms
all work with polynomial space, hence we do not mention it
explicitly.  An enumeration algorithm with polynomial delay can be
further required  to
output the elements in some order (e.g. lexicographic order) (see  \cite{jpy88}).

\bigskip

Let us now make explicit the main tool that we will use in order to get
complexity classifications.

\begin{prop}
\label{prop:circ-property}
Let $\mathbf{Prob}$ be one of the decision problems introduced above, and let
$B_1,B_2$ be finite sets of Boolean functions such that
$B_1\subseteq\pc{B_2}$. Then
$\textnormal{\textbf{Prob}}(B_1)\redl\textnormal{\textbf{Prob}}(B_2)$. In
particular, if $\pc{B_1}=\pc{B_2}$, then
$\textnormal{\textbf{Prob}}(B_1)\redleq\textnormal{\textbf{Prob}}(B_2)$.
\end{prop}

\begin{proof}
 Since all of the problems that we study
in this paper only consider the function
computed by the corresponding input circuits, it is clear that a
transformation converting a circuit into an equivalent one leaves the
properties considered in these decision problems invariant. Observe that 
 if $B_1$ and $B_2$ are finite sets of Boolean functions
such that $B_1\subseteq\pc{B_2}$, then every function from $B_1$ can be expressed as a $B_2$-circuit, its so-called $B_2$-representation. Thus
we can, in logarithmic space,
convert any $B_1$-circuit into an equivalent $B_2$-circuit in replacing every gate of the original circuit (which is a function from $B_1$) by its $B_2$-representation.  This
concludes the proof. Note that since $B_1$ and $B_2$ are not part of the input the cost of computing the $B_2$-representations of the functions of $B_1$ is a not taken into account.
\end{proof}

Since the above result shows that the complexity of the problems we study does not
depend on the particular base of a clone that we consider, we sometimes write
$\textnormal{\textbf{Prob}}(C)$ for a clone $C$. For example, we write 
$\PCSAT{\BF}$ to denote the satisfiability for a set $B$ with $\pc{B}=\BF$,
e.g., for $\PCSAT{\set{\land,\OR,\NOT}}$. Due to the above, choosing a different
base $B$ of $\BF$ results in a problem with the same complexity.

As a consequence in order to get a complete classification for
$\textnormal{\textbf{Prob}}(B)$ for every finite set $B$ it is 
enough to examine all possible clones. When we show a  hardness result for 
$\textnormal{\textbf{Prob}}(C)$ for some clone $C$,
 then hardness also holds for every finite set $B$ such that $C\subseteq
\pc{B}$. 
Also when we show tractability of $\textnormal{\textbf{Prob}}(C)$, then
tractability also holds for every finite set $B$ such that  $B\subseteq
\pc{C}$. 

We also note that a similar result as Proposition~\ref{prop:circ-property} applies
to the enumeration problem. For example, if $B_1\subseteq\pc{B_2}$, and there is a 
polynomial-delay enumeration algorithm for $B_2$-circuits, then there also is a
polynomial-delay enumeration algorithm for $B_1$-circuits.

\section{Equivalence- and isomorphism problems}
\label{EqIsoSec}
\label{sect:eqiso}
In this section we use the inclusion structure of all closed classes
(see Figure~\ref{platt}) to determine the complexity of \PCEQ{B} step
by step. Similarly we are able to give lower bounds for the
isomorphism problem of $B$-circuits. 

The following lemma will be a useful fact in our
proofs---the lemma follows from the simple observation that for the equivalence-
or isomorphism problems, consistently swapping $0$s and $1$s does not change the 
complexity. Let us first introduce some notation.
If  $f$ is an
$n$-ary Boolean function, then $\dual{f}$ denotes the Boolean function such that $\dual{f}(x_1,\dots,x_n)=\neg f(\neg x_1,\dots,\neg x_n)$. 
For a set $B$ of Boolean functions, let $\dual{B}=\set{\dual{f}\ \vert\ f\in B}$. 

\begin{lem}\label{lemma:duality}
 Let $B$ be a finite set of Boolean functions. Then $\PCEQ{B} \redleq
\PCEQ{\dual{B}}$ and $\PCISO{B} \redleq \PCISO{\dual{B}}$. 
\end{lem}

Let us first identify the tractable cases. The next proposition says that when 
besides constants only $\vee$-functions or only $\wedge$-functions or only
$\oplus$-functions are allowed, equivalence and isomorphism are easily
checkable.

\begin{prop}
\label{prop:eq_N}\label{prop:eq_EVL}
Let $B$ be a finite set of Boolean functions. If $B \subseteq \EtC$ or ${B}
\subseteq \VelC$ or ${B} \subseteq \Lin$ then
$\PCEQ{B}$ and $\PCISO{B}$ are tractable.
\end{prop}

\begin{proof}
If $B$ only contains to $\OR$-functions ($\land$-functions,
$\XOR$-functions resp.),  the
basic idea is that we can first compute an explicit normal form for the
functions computed by such a $B$-circuit. This normal form then easily allows to
determine equivalence or isomorphism.

First let $B\subseteq \VelC$. Let
$C_1(\enu{x}{1}{n})$ and $C_2(\enu{x}{1}{n})$ be two $B$-circuits. The
Boolean functions described by $C_1$ and $C_2$ can be expressed as
follows: $\boolf{C_1}{\enu{x}{1}{n}} = a_0 \vee (a_1 \wedge x_1) \vee
\dots \vee (a_n \wedge x_n)$ and $\boolf{C_2}{\enu{x}{1}{n}} = b_0
\vee (b_1 \wedge x_1) \vee \dots \vee (b_n \wedge x_n)$, where
$\enu{a}{1}{n},\enu{b}{1}{n} \in \{0,1\}$.\\
The values of $a_i$
and $b_i$, where $0 \le i \le n$, can be determined easily by using the
following simple facts: $a_0 = 0$ ($b_0 = 0$, resp.) \gdw $\boolf{C_1}{0^n} =
0$ ($\boolf{C_2}{0^n} = 0$, resp.) and $a_i = 0$ ($b_i = 0$, resp.)
for $1 \le i \le n$ \gdw $a_0=0$ ($b_0=0$, resp.)  and
$\boolf{C_1}{0^{i-1}10^{n-i}} = 0$ ($\boolf{C_2}{0^{i-1}10^{n-i}} =
0$, resp.). This can be checked in polynomial time with the help of
\PCVAL{B} as an oracle. Since $\PCVAL{B}$ is tractable (see Proposition
\ref{prop:Circval}) we conclude that the normal forms can be computed
efficiently.
Now, clearly $(C_1, C_2) \in \PCEQ{B}$ \gdw either $a_0 = b_0 = 1$ or $a_0
= b_0 = 0$ and $a_i = b_i$ for $1 \le i \le n$, and 
similarly, $(C_1, C_2) \in \PCISO{B}$ \gdw either $a_0 = b_0 = 1$ or
$|\sset{ i \mid a_i = 1, 1 \le i \le n }| = |\sset{ i \mid b_i = 1, 1
\le i \le n }|$ and $a_0 = b_0 = 0$. Thus we
conclude that $\PCEQ{B}$ and  $\PCISO{B}$ are tractable.

Tractability for  $B \subseteq \EtC$ now follows immediately
from the above using Lemma \ref{lemma:duality}.

Finally let $B \subseteq \Lin$ and let $C_1$ and
$C_2$ be $B$-circuits. The Boolean functions described by the
$B$-circuits $C_1(\enu{x}{1}{n})$ and $C_2(\enu{x}{1}{n})$ can be
expressed as follows: $\boolf{C_1}{\enu{x}{1}{n}} = a_0 \oplus (a_1
\wedge x_1) \oplus \dots \oplus (a_n \wedge x_n)$ and
$\boolf{C_2}{\enu{x}{1}{n}} = b_0 \oplus (b_1 \wedge x_1) \oplus \dots
\oplus (b_n \wedge x_n)$, where $\enu{a}{1}{n},\enu{b}{1}{n} \in
\sset{0,1}$.
Similar to the above cases the
values $a_i$ and $b_i$ for $0 \le i \le n$ can be determined by a
\ParL-calculation, since we know  that
$\PCVAL{B}$ is tractable. In particular $a_0 = \boolf{C_1}{0, \dots ,0}$,
$b_0 = \boolf{C_2}{0, \dots ,0}$, $a_i = \boolf{C_1}{0^{i-1}10^{n-i}}
\oplus a_0$ and $b_i =\boolf{C_2}{0^{i-1}10^{n-i}} \oplus b_0$, where
$1 \le i \le n$.
Now, clearly $(C_1, C_2) \in \PCEQ{B}$ \gdw $a_i = b_i$ for $0 \le i \le
n$,
and $(C_1, C_2) \in
\PCISO{B}$ \gdw $a_0 = b_0$ and $|\sset{ i \mid a_i = 1 , 1 \le i \le n }|
= |\sset{ i \mid b_i = 1 , 1 \le i \le n}|$. Again, both problems are tractable.
\end{proof}

The main step in obtaining hardness for the remaining cases now is to show that
both equivalence and isomorphism are hard for monotone functions. This is the
statement of the next lemma.

\begin{lem}
\label{lemma:EQ_and_or}
$\PCEQ{\{\OR, \land\}}$ and $\PCISO{\{\OR, \land\}}$ are \redl-hard for \coNP.
\end{lem}

\begin{proof}
We prove that \TTAUT, the problem of deciding whether a \TDNF formula is a
tautology, is logspace reducible to $\PCEQ{\{\OR, \land\}}$ and $\PCISO{\{\OR,
\land\}}$. Since  \TTAUT is well known to be \coNP-hard this will complete the
proof. 

Let $H(\enu{x}{1}{n})$ be a \TDNF formula with $\var{H}=\{x_1,\ldots , x_n\}$. 
Let $C(\enu{x}{1}{n},\enu{y}{1}{n})$ be the
circuit obtained from $H$ in replacing every occurrence of a
negated variable $\neg x_i$ by the fresh variable $y_i$.  Note that $C$ is a
$\{\OR, \land\}$-circuit that can have fictive variables. Define
$C_1(\enu{x}{1}{n},\enu{y}{1}{n})= \bigwedge_{i=1}^n (x_i\vee y_i)$ and
$C_2(\enu{x}{1}{n},\enu{y}{1}{n})=C_1\wedge C$. 
Observe that $\sat(C_2)\subseteq\sat(C_1)$.

We claim that $H$ is a tautology if and only if $C_1\loeq C_2$ if and
only if $C_1\loiso C_2$.  Suppose first that $H$ is a tautology. We
prove that every assignment that sets at least one of $x_i$ and $y_i$
to true for every $i=1,\ldots, n$ satisfies $C$, thus proving
$C_1\loeq C_2$ and \textit{a fortiori} $C_1\loiso C_2$.  Let $I$ be
such an assignment. Consider the assignment $\tilde I$ defined by
$\tilde I(x_i)=I(x_i)$ and $\tilde I(y_i)=1-I(x_i)$ for $i=1,\ldots,
n$.  Observe that $\tilde I$ satisfies $C$ since $H$ is a
tautology. Moreover since for every $i$, $I(x_i)+I(y_i)\ge 1$ we have
$I\ge \tilde I$. Therefore by monotonicity $I$ satisfies $C$ as well.

Conversely, suppose that $H$ is not a tautology. We prove that
$\#\sat(C_1)\ne\#\sat(C_2)$, thus proving  $C_1\not\loiso C_2$ and  \textit{a
fortiori}  $C_1\not\loeq C_2$. Let $I$ be an assignment that does not satisfy
$H$. Consider $I'$ defined by
$I'(x_i)=I(x_i)$ and $ I'(y_i)=1-I(x_i)$ for $i=1,\ldots, n$.
Observe that $I'$ satisfies $C_1$ but not $C_2$. Since
$\sat(C_2)\subseteq\sat(C_1)$, this proves that $\#\sat(C_2)<\#\sat(C_1)$.
\end{proof}

The next three propositions generalize the hardness result from
Lemma~\ref{lemma:EQ_and_or}.

\begin{prop}
\label{prop:Eq_S}
Let $B$ be a set of Boolean functions such that $\NRMsep \subseteq
\pc{B}$ or $\ORMsep \subseteq \pc{B}$, then $\PCEQ{B}$  and $\PCISO{B}$ are
\redl-hard for \coNP.
\end{prop} 
\begin{proof}
First let $\NRMsep \subseteq \pc{B}$. By Figure \ref{Bases} we know
that $g(x,y,z) = x \wedge (y \vee z)$ is a base of \NRMsep. Since
$g(x,y,y) = x \wedge y$ and $g(1,x,y) = x \vee y$ we know that $\land
\in \pc{B}$ and $\OR \in \pc{B \cup \{ 1 \}}$. Therefore according to
Proposition \ref{prop:circ-property} and Lemma \ref{lemma:EQ_and_or} we get
that $\PCEQ{B \cup \{ 1 \}}$  and $\PCISO{B \cup \{ 1 \}}$ are \coNP-hard. We
now reduce these problems respectively to $\PCEQ{B}$  and $\PCISO{B}$. Let
$C_1(\enu{x}{1}{n},1)$ and  $C_2(\enu{x}{1}{n},1)$ be two $B \cup \{ 1
\}$-circuits. Let $v$ be a fresh variable that will be used to replace the
constant 1. Let $C'_1(\enu{x}{1}{n},v)=C_1(\enu{x}{1}{n},v)\wedge v$ and 
$C'_2(\enu{x}{1}{n},v)=C_2(\enu{x}{1}{n},v)\wedge v$. Since $\land
\in \pc{B}$, $C'_1$ and $C'_2$ can be represented as $B$-circuits, and their
$B$-representation can be computed in logarithmic space, see proof of
Proposition~\ref{prop:circ-property}.  It is obvious that 
$C_1\loeq
C_2$ if and only if  $C'_1\loeq C'_2$. If  $C_1\loiso C_2$ then clearly 
$C'_1\loiso C'_2$. Conversely if $C'_1\loiso C'_2$ then there is a permutation
$\pi\colon\{x_1,\ldots x_n, v\}\rightarrow\{x_1,\ldots x_n, v\}$ such that for
all assignment $I$, $I\models C'_1$ if and only if $\pi(I)\models C'_2$. Since
the value of $v$ is fixed to $1$ in every satisfying assignment one can suppose
w.l.o.g. that $\pi(v)=v$. In this case we clearly have that for
all assignment $I$, $I\models C_1$ if and only if $\pi(I)\models C_2$, thus
showing that $C_1\loiso C_2$ .

Now let $\ORMsep \subseteq \pc{B}$. By inspecting Figure \ref{platt} we obtain
that $\NRMsep \subseteq \pc{B}$ as well, or $\NRMsep \subseteq
\dual{\pc{B}}$. According to Lemma \ref{lemma:duality} the
proof is then completed.
\end{proof}

\begin{prop}
\label{prop:Eq_D}
Let $B$ be a finite set of Boolean functions such that $\Tself
\subseteq \pc{B} \subseteq \Self$, then \PCEQ{B} is \redl-hard for $\coNP$.
\end{prop}

\begin{proof}
Due to Lemma~\ref{lemma:EQ_and_or}, we know that $\PCEQ{\set{\land, \OR}}$ is \redl-hard
for $\coNP$. Hence let $C_1$ and $C_2$ be $\set{\land,\OR}$-circuits, where 
$\var{C_1}\cup\var{C_2}=\set{x_1,\dots,x_n}$.
By Figure \ref{Bases} we know that $t_2(x,y,z)
= (x \land y) \lor(x \land z) \lor (y \land z)$, the ternary majority function,
is a base for
$\Tself$.
Using the equalities
$t_2(x,y,0) = x \land y$ and $t_2(x,y,1) = x \lor y$, we can transform $C_1$ and $C_2$
into equivalent $\set{t_2,0,1}$-circuits in logarithmic space. For ease of notation,
we denote these (equivalent) circuits with $C_1(\enu{x}{1}{n},0,1)$ 
and $C_2(\enu{x}{1}{n},0,1)$ again. We note that due to the above transformation,
every application of $t_2$ in $C_1$ or $C_2$ has exactly one constant argument.

We now construct $\set{t_2}$-circuits $C_1'$ and $C_2'$ such that $C_1\loeq C_2$ if and only
if $C_1'\loeq C_2'$. Due to Proposition~\ref{prop:circ-property}, this completes the proof.

Let $u$ and $v$ be  fresh variables that will be used to replace the
constants 0 and 1 that appear in $C_1$ and $C_2$. Now define

\noindent
$C'_1(\enu{x}{1}{n},u,v)=t_2(v, C_1(\enu{x}{1}{n},u,v), u)$ and \\
$C'_2(\enu{x}{1}{n},u, v)=t_2(v, C_2(\enu{x}{1}{n},u,v), u)$. 

By construction, $C'_1$ and $C'_2$ are $\set{t_2}$-circuits.
We prove that $C_1\loeq C_2$ if and only if $C'_1\loeq C'_2$. 

First assume that $C_1\loeq C_2$, and let $I$ be an assignment for $\set{\enu x1n,u,v}$. If $I(u)=I(v)$, then
$I\models C_1'$ iff $I\models C_2'$ iff $I(u)=I(v)=1$. Now consider the case that $I(u)\neq I(v)$.
Since $t_2$ is a self-dual function it is sufficient to consider the case $I(u)=0$ and
$I(v)=1$. In this case $I\models C_i'$ iff $I\models C_i$ for $i=1, 2$, and  hence $I\models C_1'$ iff $I\models C_2'$ since $C_1\loeq C_2$.

Conversely, suppose that $C_1\not\loeq
C_2$. Then we can suppose that there exists an assignment $I$  that satisfies
$C_1$ but not $C_2$.  Extend $I$ to $I'$ by setting $I'(u)=0$ and $I'(v)=1$. 
It is easy to see that $I'$ satisfies $C'_1$ but not $C'_2$, thus
 proving that  $C'_1\not\loeq C'_2$.
\end{proof}

We have a similar result for the isomorphism problem. 

\begin{prop}
\label{prop:ISO_D}
Let $B$ be a finite set of Boolean functions such that $\Tself
\subseteq \pc{B} \subseteq \Self$,  then \PCISO{B} is \redl-hard
for
$\coNP$.
\end{prop}

While the proof of this proposition uses essentially the same reduction as above, it is technically more involved 
and requires some technical results. We will reduce from the isomorphism problem for $\set{\land,\OR}$-circuits,
which we know to be $\coNP$-hard due to Lemma~\ref{lemma:EQ_and_or}. However, in the proof of Proposition~\ref{prop:ISO_D},
we will need some special properties of the instances of $\PCISO{\{\OR, \land\}}$ that we reduce from. We therefore
present a series of intermediate technical results that allow us to restrict the instances of $\PCISO{\{\OR, \land\}}$
as required.

First we need to introduce some new notion.

\begin{defi}
   A pair of two variables $\{s,t\}$ is \emph{dominant} for a
    circuit $C$, if every truth assignment $I$ with $I(s)=I(t)=\alpha$ 
satisfies the circuit $C$ if and only if $\alpha=1$.
\end{defi}

The following lemma gives some easy properties of dominant pairs. 
The proof of the lemma is straight-forward.

\begin{lem}\label{lemma:dominant pairs not disjoint}\hfill
\begin{enumerate}[\em(1)]
 \item  Let $C$ be a circuit, and let $\{s, t\}$ and $\{s', t'\}$ be two
dominant pairs for $C$. 
Then $\{s, t\}\cap \{s', t'\}\neq\emptyset$.
\item  Let $C_1$ and $C_2$ be two circuits such that  $C_1\cong C_2$ via a
permutation $\pi$.
 If $\{s, t\}$ is
 a dominant pair for $C_1$, then $\{\pi(s), \pi(t)\}$ is a dominant pair for
$C_2$.
\end{enumerate}
\end{lem}

\begin{proof}
For the first part, assume that $\{s,t\}\cap\{s',t'\}=\emptyset$. Then there is an assignment $I$
with $I(s)=I(t)=0$, and $I(s')=I(t')=1$. Since $\{s,t\}$ is dominant for $C$, it follows that $I\nmodels
C$. On the other hand, since $\{s',t'\}$ is dominant for $C$ as well, we
know that $I\models C$, a contradiction.
The second part is trivial.
\end{proof}

 Next we need the following result, which says that the isomorphism problem
remains hard for monotone functions even for some restricted instances.

\begin{lem}
 \label{lemma:ISO_and_or_restr}
$\PCISO{\{\OR, \land\}}$ is \redl-hard for \coNP,  even when instances are
restricted to pairs of circuits
 $(C_1, C_2)$ where  neither $C_1$ nor $C_2$ implies or is implied by one
variable, and further
 $\#\sat(C_1)+\#\sat(C_2) <2^n$ where $n$ is the number of variables in $C_1$,
which is the same as the number of variables in $C_2$.
\end{lem}

\begin{proof} We reduce the problem $\PCISO{\{\OR, \land\}}$,  which is \coNP-hard due to
 Lemma~\ref{lemma:EQ_and_or}, to the same problem with restrictions on the instances.

 Let $(C_1^0, C_2^0)$ be a pair of $\{\OR, \land\}$-circuits given as an instance of
 $\PCISO{\{\OR, \land\}}$.

Without loss of generality one can suppose that they have the same number of
variables, $n$, and that
$C_i^0$ is of the form
${C_i^0}'\wedge (y\wedge z)$. Thus $C_i^0$ has fewer than $2^{n-1}$
solutions. Since both $C_i^0$ are monotone, we
can, in polynomial time, verify whether $C_1^0$ or $C_2^0$ are constant. If
one of them is, then $C_1^0\cong C_2^0$ is true if and only if they are
equivalent to the same constant. Hence we assume that neither $C_1^0$
nor $C_2^0$ is constant.
For
$i\in\set{1,2}$, we now rewrite $C_i^0$ into

$$C_i=t_2(z_1,z_2,C_i^0),$$ where $t_2$ is the ternary majority function.

 We will show that the $C_i$'s have the desired properties, and that $C_1^0\cong C_2^0$
if and only if $C_1\cong C_2$, thus concluding the proof.

 Clearly, since the outmost operator of $C_1$ and $C_2$ is the majority
 function, and neither $C_1^0$ nor $C_2^0$ are constant, it follows that
 no variable implies or is implied by one of the $C_i$.
 Obviously, $\#\sat(C_i)=2\cdot\#\sat(C_i^0)+ 2^n$, since the solutions of $C_i$
 are exactly those of $C_i^0$ extended with $z_1\neq z_2$ (giving two
 solutions for each solution of $C_i^0$), plus all $2^n$ assignments
 setting $z_1=z_2=1$. Since $\#\sat(C_i^0)<2^{n-1}$, it follows that
 $\#\sat(C_i)<2^n+2^n=2^{n+1}$. Therefore, $\#\sat(C_1)+\#\sat(C_2)<2^{n+2}$ as
 required (note that $n+2$ is the number of variables appearing in
 $C_1$ and $C_2$).

 It remains to show that 
 $C_1^0\cong C_2^0$ if and only if $C_1\cong C_2$. The left-to-right
 direction is trivial, by extending the permutation to be the identity
 on $\set{z_1,z_2}$. For the other direction, assume that $C_1\cong
 C_2$ via a permutation $\pi$. Observe that $\set{z_1,z_2}$ is a dominant pair for
 both circuits. If
 $\pi(\set{z_1,z_2})=\set{z_1,z_2}$, then since $z_1$ and $z_2$ are
 symmetric, we can assume that $\pi(z_i)=z_i$, and $\pi$ restricted to
 the original variables establishes $C_1^0\cong C_2^0$.

 Hence assume $\pi(\set{z_1,z_2})\neq\set{z_1,z_2}$. According to Lemma
\ref{lemma:dominant pairs not disjoint},
 $\pi(\set{z_1,z_2})\cap\set{z_1,z_2}\neq\emptyset$. Since $z_1$ and
 $z_2$ are symmetric, we assume without loss of generality that
 $\pi(\set{z_1,z_2})=\set{z_1,x}$ for a variable $x$ of $C_2$.

 We prove that $C_2^0$ is equivalent to $x$. First let $I$ be an
 assignment to the variables in $C_2^0$ with $I(x)=1$, we prove that
 $I\models C_2^0$. For this, consider the assignment $I^+$ which extends
 $I$ by $I^+(z_1)=1$ and $I^+(z_2)=0$. Since $\set{z_1,x}$ is
 dominant, it follows that $I^+\models C_2$, and thus $I\models
 C_2^0$. Therefore, $x$ implies $C_2^0$. For the other direction, let $I$
 be an assignment with $I(x)=0$, we show that $I\nmodels C_2^0$. We
 extend $I$ to $I^+$ by setting $I^+(z_1)=0$ and $I^+(z_2)=1$. Since
 $\set{z_1,x}$ is dominant for $C_2$, it follows that $I^+\nmodels
 C_2$, hence we know that $I\nmodels C_2^0$, and thus $C_2^0$ is
 equivalent to $x$ as claimed.

 Therefore, $C_2$ has exactly three relevant variables (recall that a variable $x$ is \emph{relevant} for a circuit $C$, if there
 are assignments $I_1$ and $I_2$ such that $I_1(x')=I_2(x')$ for all variables $x'\neq x$, and $I_1\models C$ and $I_2\nmodels C$,
 i.e., if the value of the function computed by the circuit in fact depends on $x$).

 Since
 $C_2\cong C_1$, we know that $C_1$ also has exactly three
 relevant variables. Hence $C_1^0$ has exactly one relevant variable,
 and since we also know that $C_1^0$ is a monotone circuit, it follows that $C_1^0$ is 
 equivalent to a single variable. In particular, $C_1^0\cong
 C_2^0$ as claimed.
\end{proof}

We need a last technical result. This lemma allows us, in the later proof of Proposition~\ref{prop:ISO_D},
to use a similar argument as in the above proof of Lemma~\ref{lemma:ISO_and_or_restr}: In both proofs it
is essential that we can control the possible dominant pairs of a circuit that is of the form
$t_2(x_1,x_2,C)$, where $t_2$ is the ternary majority function and 
$C$ is some circuit. In the proof of Lemma~\ref{lemma:ISO_and_or_restr}, we knew the dominant sets
of $t_2(z_1,z_2,C_i^0)$ since $z_1$ and $z_2$ did not appear in $C_i^0$. In the proof of Proposition~\ref{prop:ISO_D},
the situation will be a bit more complicated, and we will need the following lemma to ensure that
the dominant pairs in the circuits resulting from our reduction are exactly the ones that we need.

\begin{lem}\label{lemma:unique dominant pair}
 If $C(x_1,\dots,x_n,1,0)$ is a circuit that does not imply a
 variable and is not implied by a variable, then $\set{u,v}$ is the
 only dominant pair for
$$C'=t_2(u,C_2(x_1,\dots,x_n,u,v),v),$$ where $u$ and $v$
 are new variables.
\end{lem}

We note that a circuit $C$ implies a variable if and only if the function computed by $C$ is $1$-separating, and
is implied by a variable if and only if the function computed by $C$ is $0$-separating
We note that for the ternary majority function $t_2$, all pairs of two distinct variables are dominant.

\begin{proof}
 We use the following notation: For an assignment $I$ for $C$, with
 $I^+$ we denote the assignment $I$ extended with $I^+(u)=1$ and
 $I^+(v)=0$. By construction it follows that $I\models C$ if and
 only if $I^+\models C'$.

 Clearly, $\set{u,v}$ is dominant for $C'$. Let
 $\set{u',v'}\neq\set{u,v}$ be dominant for $C'$. Due to
 Lemma~\ref{lemma:dominant pairs not disjoint}, we know
 that $\set{u,v}\cap\set{u',v'}\neq\emptyset$.

 First assume $u\in\set{u,v}\cap\set{u',v'}$, then
 $\set{u',v'}=\set{u,x}$ for a variable $x$ of $C$. We prove that
 $x$ implies $C$. Hence let $I(x)=1$. By construction, we have that
 $I^+(u)=1$, and $I^+(x)=I(x)=1$. Since $\set{u,x}$ dominates $C'$,
 it follows that $I^+\models C'$. Due to the above, this means that
 $I\models C$. Hence $x$ implies $C$, a contradiction.

 Similarly, assume that $v\in\set{u,v}\cap\set{u',v'}$, then
 $\set{u',v'}=\set{v,x}$ for a variable $x$ of $C$. We claim that
 $C$ implies $x$. Hence let $I$ be an assignment with $I\models
 C$, and assume that $I(x)=0$. From the above it follows
 that $I^+\models C'$. On the other hand, we have that
 $I^+(x)=I^+(v)=0$, and since $\set{v,x}$ dominates $C'$, this
 implies $I^+\nmodels C'$, a contradiction. Therefore, $C$ indeed
 implies $x$, which is a contradiction to the prerequisites of the
 lemma.
\end{proof}
 
  We are now in a position to prove Proposition \ref{prop:ISO_D}.

\begin{proof}
As in the proof of Proposition \ref{prop:Eq_D} we get
that $\PCISO{B \cup \{ 0,1 \}}$ is \coNP-hard,
and we reduce this problem  to $\PCISO{B}$.  
Let
$C_1(\enu{x}{1}{n},0,1)$ and  $C_2(\enu{x}{1}{n},0, 1)$ be two $B \cup \{0, 1
\}$-circuits. According to Lemma \ref{lemma:ISO_and_or_restr} one can suppose
that 
neither $C_1$ nor $C_2$ implies or is implied by one variable, and further that
 $\#\sat(C_1)+\#\sat(C_2) <2^n$ where $n$ is the number of variables in $C_1$,
which is the same as the number of variables in $C_2$.
 
Let $u$ and $v$ be fresh variables, and let
$$C'_1(\enu{x}{1}{n},u,v)=t_2(u, C_1(\enu{x}{1}{n},u,v), v)$$
 and  
$$C'_2(\enu{x}{1}{n},u, v)=t_2(u, C_2(\enu{x}{1}{n},u,v), v).$$

As mentioned in the earlier proof of Lemma~\ref{lemma:ISO_and_or_restr}, this construction is very similar
to what we used there. The major difference lies in the role of the variables that are used in the
application of the newly introduced majority function: In the proof of Lemma~\ref{lemma:ISO_and_or_restr},
we used new variables $z_1$ and $z_2$ that did not appear anywhere else, and whose role was symmetric. In fact,
the proof of Lemma~\ref{lemma:ISO_and_or_restr} only works since $z_1$ and $z_2$ did not appear in the circuits
$C_i^0$ considered in that proof.

In the current proof, the situation is different: Here, the variables $u$ and $v$ do appear in the circuits
$C_i(\enu{x}{1}{n},u,v)$, and they are clearly not symmetric---they ``simulate'' the values $0$ and $1$, respectively.
In the remainder of the current proof, we make crucial use of the facts established in Lemma~\ref{lemma:ISO_and_or_restr},
namely, that the circuits $C_1$ and $C_2$ are not implied by, or imply, a variable. This then allows us to apply
Lemma~\ref{lemma:unique dominant pair} and ensure that $\set{u,v}$ is the only dominant pair of $C'_i$.

Another difference is that in the current proof, the circuits $C_i(\enu{x}{1}{n},u,v)$ are indeed $\Tself$-circuits,
where in the earlier result, the majority function was applied to (almost) arbitrary $\set{\wedge,\vee}$-circuits.

We   prove  
 $C_1\loiso
C_2$ if and only if  $C'_1\loiso C'_2$. It is obvious that if  $C_1\loiso
C_2$ then  $C'_1\loiso C'_2$. Conversely, suppose that  $C'_1\loiso C'_2$. Then
there exists a permutation   $\pi \colon \sset{\range{x_1}{x_n}, u,v}
\rightarrow \sset{\range{x_1}{x_n}, u,v}$
such that for every truth assignment   $I\colon
\sset{\enu{x}{1}{n}, u,v }\rightarrow \{0,1\}$ it holds that $I\models C'_1$ if
and only if $\pi(I)\models C'_2$. Observe that because of the majority function,
 the pair
$\{u,v\}$ is  dominant   for both circuits $ C'_1$ and $ C'_2$. According to
Lemma \ref{lemma:unique dominant pair}
we have then $\{\pi(u),
\pi(v)\}=\{u,v\}$. Suppose that $\pi(u)=v$ and
$\pi(v)=u$. Let $\#C_1'$ be the number of truth assignments satisfying
$C'_1$ that set $u$ to $0$ and $v$ to 1, and $\#C_2'$ be the number of
truth assignments satisfying $C'_2$ that set $u$ to $1$ and $v$ to
0. Since $C'_1$ and $C'_2$ are isomorphic through a permutation $\pi$
such that $\pi(u)=v$ and $\pi(v)=u$ we have $\#C_1'=\#C_2'$.  We have
$\#C_1'=\#\{I \mid I(u)=0, I(v)=1 \hbox{ and } C_1(I(x_1),\ldots ,
I(x_n), 0, 1)=1\}$ and $\#C_2'=\#\{J \mid J(u)=1, J(v)=0 \hbox { and }
C_2(J(x_1),\ldots , J(x_n), 1, 0)=1\}$. Observe that
$\#C_1'= \#\sat(C_1)$, while
\[\#C_2'= \#\{I\colon
\sset{\enu{x}{1}{n}}\rightarrow \{0,1\}\mid C_2(I(x_1),\ldots , I(x_n),
1, 0)=1\}$$
$$ = \#\{I\colon
\sset{\enu{x}{1}{n}}\rightarrow \{0,1\}\mid C_2(1-I(x_1),\ldots , 1-I(x_n),
0, 1)=0\}, \] 
since $C_2$ is a $B \cup \{0, 1
\}$-circuit and $B$ contains only self-dual
functions. Therefore $\#C_2'=2^n-\#\sat(C_2)$. But $\#C_1'=\#C_2'$ implies $
\#\sat(C_1)= 2^n-\#\sat(C_2)$, \textit{i.e.}, $\#\sat(C_1)+\#\sat(C_2)=2^n$,
which is not the case by assumption, thus providing a contradiction. Therefore 
$\pi(u)=u$ and $\pi(v)=v$. With this it is easy to see that  $C'_1\loiso C'_2$
implies that  $C_1\loiso C_2$ through the same permutation $\pi$.
\end{proof}

By a careful inspection of Figure \ref{platt} we see that
Propositions~\ref{prop:eq_EVL}, \ref{prop:Eq_S}, \ref{prop:Eq_D}, and \ref{prop:ISO_D} cover all
cases.
This leads us to the following classification theorems for the
complexity of the equivalence- and isomorphism- problems of
$B$-circuits:

\begin{thm}
\label{ISOEQDich}
Let $B$ be a finite set of Boolean functions. 
\begin{enumerate}[\em(1)]
\item  If $B \subseteq \EtC$ or $B \subseteq \VelC$ or $B
\subseteq \Lin$ then  \PCEQ{B} and \PCISO{B} are tractable.
\item In all other cases $\PCEQ{B}$ is $\redl$-complete for $\coNP$ and
$\PCISO{B}$ is
$\redl$-hard for $\coNP$.
\end{enumerate}
\end{thm} 

\section{Results for audit-like problems}
\label{sect:audit}

This section covers our results about the problems related to
the audit and frozen variable problems. We start with the following
basic facts about  complexity upper bounds.

\begin{prop}
For every finite set $B$ of Boolean functions, the following upper bounds hold:
\begin{multicols}{2}
\raggedcolumns
\begin{enumerate}[\em(1)]
\item $\FV {B} \in \DP$,
\item $\EFV {B} \in \DP$,
\item $\USAT{B} \in \DP$, and
\item $\AUDIT{B} \in \coNP$.
\end{enumerate}
\end{multicols}
\end{prop}

As an auxiliary problem we will first examine the generalization of the
satisfiability problem  $\SATstar{B}$, which asks whether a $B$-circuit has
a satisfying assignment different from the all 1's one. This problem was
examined in \cite{crhe97} in the constraint setting.

\begin{thm}
\label{thm:selectSAT}
Let $B$ be a finite set of Boolean functions. Then   $\SATstar{B}$ is
$\NP$-complete if  $\NRsep \subseteq \pc{B}$, and solvable in  $\P$ otherwise.
\end{thm}

\begin{proof}
First assume $\NRsep \subseteq \pc{B}$. In this case 
by looking at Post's lattice (see Figure \ref{platt}) we know that
$[\NRsep\cup\{0\}]=\Nsep$, hence following Proposition \ref{prop:Csat},
$\PCSAT{\NRsep\cup\{0\}}$ is $\NP$-complete. We will now
reduce $\PCSAT{\NRsep\cup\{0\}}$ to $\SATstar{\NRsep}$. Given an
$(\NRsep\cup\{0\})$-circuit $C(\enu{x}{1}{n},0)$ we use a new variable
$x$ as a replacement for the constant $0$. Thus we obtain an
$\NRsep$-circuit $C'(\enu{x}{1}{n},x)$. Looking at Table 1 we see that
the Boolean function $g(x,y,z)=x\wedge (y\vee\bar z)$ belongs to
$\NRsep$. Hence, let $\hat{C}$ be the $\NRsep$-circuit defined by 
$\hat{C}=g(\cdots g(g(C', x_1,x), x_2,x),\cdots, x_n,x)$. 
Observe that $\hat{C}$ 
is equivalent to 
 $C'\wedge(x_1\vee\bar x) \wedge (x_2\vee\bar
x)\wedge \cdots \wedge(x_n\vee\bar x).$
Observe now that $C$ has a satisfying assignment if and only if there
is an assignment different from the all $1$'s one that satisfies
$\hat{C}$. We conclude that $\SATstar{\NRsep}$ is $\NP$-hard, thus showing that
$\SATstar{B}$ is
$\NP$-complete for all $B$ such that  $\NRsep \subseteq \pc{B}$.

If $B \subseteq M$, then an $n$-ary circuit $C$ has a satisfying assignment besides the all-$1$-assignment
 if and only if it has a satisfying assignment of the form $1^i01^{n-i-1}$.
Hence $\SATstar{B}$ can be solved with $n$ evaluations of the circuit $C$.

If $B \subseteq \Lin$, we can use the linear normal form $\czero\oplus
\bigoplus_{i=1}^n c_i x_i$, which is obviously polynomial time
computable, to solve $\SATstar{B}$ efficiently.  

If $B \subseteq \Self$
or $B \subseteq \Omsep{2}$ then we claim that  each $n$-ary $B$-circuit has at
least
$2^{n-1}$ satisfying assignments, which obviously makes
$\SATstar{B}$ tractable. If  $B \subseteq \Self$ the claim holds for any self-dual
circuit has exactly $2^{n-1}$ solutions. If  $B \subseteq \Omsep{2}$, 
note that for every $B$-circuit and compatible assignment $I$, if $I$ does not
satisfy $C$, then $\dual I$ does. Indeed, assume that both $I$ and $\dual{I}$ do not
satisfy $C$. Since $C$ is a $B$-circuit, and $B\subseteq\Omsep 2$, we
 know that the function described by $C$ is $0$-separating of degree
 $2$. Thus every set $S$ with $\card{S}=2$ and $S\subseteq
 C^{-1}(\set{0})$ is $0$-separating. The set $S$ defined as
 $S\defeq\set{(I(x_1),\dots,I(x_n)),(\dual{I}(x_1),\dots,\dual{I}(x_n))}$
 meets these conditions, and hence is $0$-separating. From the
 definition, it follows that there is some $i\in\set{1,\dots,n}$ such
 that $I(x_i)=\dual{I}(x_i)$, which is a contradiction to the
 definition of $\dual{I}$. 
In particular, the number of solutions of such a circuit is at least 
$2^{n-1}$.
\end{proof}

Next we want to study problems which are related to the concept of
frozen variables. 

\begin{lem} \label{lemma:coNP_cases_FV}
Let $B$ be a finite set of Boolean functions.
If $\Nself \subseteq \pc{B} \subseteq \Self$ or $\ORsep \subseteq
\pc{B} \subseteq \Nrep$, then $\FV{B}$ and $\EFV{B}$ are
$\coNP$-complete.
\end{lem}

\begin{proof}
Observe that for all $B$ that satisfy the conditions above all
$B$-circuits are trivially satisfiable, hence $\FV{B}$ and $\EFV{B}$
are in $\coNP$.

Let $\ORsep \subseteq \pc{B} \subseteq \Nrep$. We will reduce the
$\coNP$-complete problem (see Proposition \ref{prop:Csat})
$\overline{\PCSAT{\Orep}}$ to $\EFV{B}$ and $\FV{B}$. Recall that due to
the very end of Section~\ref{subsec:Post}, we know that 
$\sset{x \vee (y \wedge \overline{z}),0}$ is a base of $\Orep$.
Hence let $C$ be a circuit over $\sset{x \vee (y \wedge \overline{z}),0}$.
We build a new circuit $C'$ out of $C$
by taking a fresh variable $x$  and by replacing
every occurrence of $0$ in $C$ with $x$. Then $C'$ is an $\ORsep$-circuit. 
Since  $\cV_2\subseteq\cS_{00}\subseteq\cS_{02}$ we have $\OR \in \pc{B}$ (see Figure \ref{platt}), and hence  
 $C' \vee x$  can be converted into an equivalent
$B$-circuit
Note that the only possibly frozen variable in $C' \vee x$ is $x$, 
since with setting $x$ to true, every possible assignment to the other
variables satisfies the circuit. Finally,
$x$ is a frozen variable in $C' \vee x$ if and only if $C$ is not
satisfiable.

Now let $\Nself \subseteq \pc{B} \subseteq \Self$.  We will reduce the
$\coNP$-complete problem $\PCEQ{\Nself}$ (see Theorem \ref{ISOEQDich})
to $\EFV{B}$ and $\FV{B}$.  For that let $C_1$ and $C_2$ be two
$n$-ary $\Nself$-circuits. Let $x$ be a fresh variable and let $C' \define x \oplus C_1 \oplus C_2$.
Since $x \oplus y \oplus z$ is a function in $\Nself$ (see Figure
\ref{platt}), $C'$ is a $\Nself$-circuit. Consider now the $\Nself$-circuit $C''(x,y,z) = t_2(x,y,z)$ (remind that $t_2\in\Nself$). The
reduction $\Phi$ works
as follows:
\begin{displaymath}
  \Phi(C_1,C_2) \define \caseDistinction{
    C'' & \mbox{, if }C_1(0^n) \neq C_2(0^n) \text{ or }C_1(1^n) \neq C_2(1^n)\\
    C'  & \mbox{otherwise}
  }
\end{displaymath}                        
We claim that $C_1 \equiv C_2$ holds if and only if $\Phi(C_1,C_2) \in
\EFV{\Nself}$ if and only if $(\Phi(C_1,C_2),\allowbreak \sset{x}) \in
\FV{B}$.  For that let $C_1 \equiv C_2$, then $\Phi(C_1,C_2)=C'$. Since
$C_1 \oplus C_2 \equiv 0$ the formula $C'$ is satisfied if and only if
$x$ is satisfied. Thus $x$ is a frozen variable and therefore
$\Phi(C_1,C_2)\in \EFV{B}$ and $(\Phi(C_1,C_2),\sset{x}) \in \FV{B}$.  On
the other hand, if $C_1 \not \equiv C_2$ then we have two cases. If
$C_1(0^n) \neq C_2(0^n)$ or $C_1(1^n) \neq C_2(1^n)$ then
$\Phi(C_1,C_2)=C''$, which is a circuit without frozen variables. If
$C_1(0^n) = C_2(0^n)$ and $C_1(1^n) = C_2(1^n)$ then $\Phi(C_1,C_2)=C'$.
Since $C_1 \not \equiv C_2$, there is an assignment $\alpha \in
\sset{0,1}^n$ such that $C_1(\alpha) \neq C_2(\alpha)$.  Therefore $1=
0 \oplus C_1(\alpha) \oplus C_2(\alpha) = 1 \oplus C_1(0^n) \oplus
C_2(0^n)= 1 \oplus C_1(1^n) \oplus C_2(1^n)$ and there is no frozen
variable in $C'$. 
\end{proof}

The following is our main classification result for the problem that asks if
there is any frozen variable:

\begin{thm}
\label{thm:existsFV}
Let $B$ be a finite set of Boolean functions.
\begin{enumerate}[\em(1)]
 \item If $B\subseteq\Lin$, $B \subseteq \Mon$, or $B=\NRsep$ then
    $\EFV{B}$ is tractable.
 \item If $\pc{B}=\Nsep$, then $\EFV{B}$ is $\NP$-complete.
 \item If $\Nsep \subset \pc{B}$ then $\EFV{B}$ is $\DP$-complete.
 \item In all other cases $\EFV{B}$ is $\coNP$-complete.
\end{enumerate}
\end{thm}

\proof\hfill
\begin{enumerate}[(1)]
\item \label{existFrozenThmTractCase} If $B \subseteq \Lin$ then in a
  $B$-circuit there is a frozen variable if and only if exactly one of
  the variables of its linear normal form has a coefficient of 1.
  Note that the linear normal form can easily be computed from the
  circuit using simulation (see proof of Proposition \ref{prop:eq_EVL}).
      
  Now let $B \subseteq \Mon$ and let $C(x_1,\dots,x_n)$ be a
  $B$-circuit. By monotonicity the variable $x_i$ is frozen if and only if $C(1^n)=1$
  and $C(1^{i-1}01^{n-i})=0$.
 Moreover the
  satisfiability of a monotonic $C$ can be easily tested. If $B=\NRsep
  = \Nrep \cap \Nsep$ then every $B$-circuit $C$ is satisfiable and
  has a frozen variable because $C$ is $1$-separating.
\item Note that an $\Nsep$-circuit has a frozen variable by
  definition if and only if it is satisfiable and it is therefore
  equivalent to $\PCSAT{B}$, which is known to be $\NP$-complete
  by Proposition \ref{prop:Csat}.
\item \label{EFVDPcompleteCases} Let $B$ such that $\Nsep \subset
  \pc{B}$. It is obvious, that $\EFV{B}$ is in $\DP$.  Let us now introduce the problem $\SATP(B)$:
\begin{eqnarray*}
   \SATP(B) \define \sset{(C_1,C_2) &\suchthat& C_1\mbox{ and
     }C_2\mbox{ are } B\text{-circuits}\\ &&\text{and }(C_1 \in
     \PCSAT{B}\text{ xor }C_2 \in \PCSAT{B})}
  \end{eqnarray*}
By definition $\SATP(B)$ is in $\DP$ and is $\DP$-complete as far as $\PCSAT{B}$ is $\NP$-complete
(cf.~\cite{CGH88,CGH89}).
 We now reduce $\SATP(B)$ to $\EFV{B}$, thus completing the proof.
 Since $\Nsep \subset \pc{B}$, there
  is a $k \geq 2$ such that $t_k$ is in $\pc{B}$.   Let $C_1$  and $C_2$ be $B$-circuits
 which are $m$- and $n$-ary
  respectively. Now define $C\define t_k(C_1,C_2,x_1,\dots,x_{k-1}),$ where
  $x_i$ is a fresh variable for $1
  \leq i \leq k-1$. Clearly $C$ is a $B$-circuit. Next we show that
  this transformation gives the needed reduction.

  If $C_1, C_2$ are both satisfiable   then there are
  assignments $\alpha \in \sset{0,1}^m$ and $\beta \in \sset{0,1}^n$
  such that $C_1(\alpha)=C_2(\beta)=1$. Then none of the variables of
  $C_1$ is frozen in $C$, since $C(\gamma\beta 1^{k-1})=1$ for all $\gamma
  \in \sset{0,1}^m$. The same argumentation holds for all variables in
  $C_2$. Furthermore for all $1 \le i \le k - 1$ it holds that
  $x_i$ is not frozen in $C$, since $C(\alpha \beta
  1^{i-1}01^{k-i-1})=C(\alpha \beta 1^{k-1})=1$.

  If  $C_1, C_2 $ are both unsatisfiable, then $C$ is
  not satisfiable and therefore has no frozen variables.

Suppose now without loss of generality $C_1$ is satisfiable and that $C_2$ is not,
 then $C \equiv C_1 \wedge x_1 \wedge
  \dots \wedge x_{k-1}$ and obviously at least all of the $x_i$'s ($1
  \leq i \leq k-1$) are frozen.
\item If $\Nself \subseteq \pc{B} \subseteq \Self$ or $\ORsep
  \subseteq \pc{B} \subseteq \Nrep$, then $\EFV{B}$ is
  $\coNP$-complete because of Lemma~\ref{lemma:coNP_cases_FV}. The
  only remaining case is $\NRsep \subset \pc{B} \subseteq \Rep$. Then
  $B$-circuits are trivially satisfiable and therefore
  $\EFV{B} \in \coNP$.  The proof of the lower bound is similar to
  Case \ref{EFVDPcompleteCases}, but this time the reduction starts
  with $\overline{\SATstar{B}}$, which is $\coNP$-complete (see
  Theorem~\ref{thm:selectSAT}). Since $\NRsep \subset \pc{B}\subseteq \Nrep$ there is a
  $k \geq 2$ such that $t_k$ is in
  $\pc{B}$.  Let
  $C(x_1,\dots ,x_n)$ be a $B$-circuit with the variables
  $x_1,\dots,x_n$. Let $C'\define
  t_k(C,y_1,\dots,y_k)$, $C'' \define \left((\bigwedge_{i=1}^k
  y_i) \vee \neg ( \bigwedge_{j=1}^n
  x_j)\right)$ and $G(x_1,\dots,x_n,\allowbreak y_1,\dots,y_k) \define
  C' \wedge C''$. Observe that $C''$ and therefore $G$ can be
  converted into equivalent $B$-circuits, because
  $\land \in \pc{B}$ and $[\sset{x \wedge
  (y \vee \overline{z})}] = \NRsep$.

 If $C$ is unsatisfiable, then because of $C'$ then $G$ is
  satisfiable only by setting $y_i$ to 1 for all $1 \leq i \leq k$.
The same holds if $C$   has the all-1 assignment as only satisfying assignment   because of $C''$.
  Hence in both cases all $y_i$ are frozen.

 On the other hand, if there is an $\alpha \in
  \sset{0,1}^n$ such that $\alpha \neq (1,\dots,1)$ and $C(\alpha)=1$
  then none of the $x_i$'s is frozen ($1 \leq i \leq n$), since $G$
  can be satisfied by just setting all the $y_i$'s to 1. Furthermore,
  for each $j \in \sset{1,\dots,k}$ holds $G(\alpha
  1^{j-1}01^{k-j})=1=G(\alpha 1^k)$, hence none of the $y_j$'s is
  frozen.\qed
\end{enumerate}

\noindent Concerning the variant of the above problem, where the
frozen variable is part of the input, we obtain the following
classification:

\begin{thm}
\label{thm:FV}
Let $B$ be a finite set of Boolean functions.
\begin{enumerate}[\em(1)]
 \item If $B\subseteq\Mon$ or $B \subseteq \Lin$, then $\FV{B}$ is
  tractable,
 \item else if $\Nsep\subseteq \pc{B}$, then $\FV{B}$ is
  $\DP$-complete,
 \item else $\FV{B}$ is $\coNP$-complete.
\end{enumerate}
\end{thm}

\begin{proof}
If $B \subseteq \Mon$ or $B \subseteq \Lin$, then the argumentations
from Theorem \ref{thm:existsFV}.\ref{existFrozenThmTractCase} hold.

We have seen in Lemma~\ref{lemma:coNP_cases_FV} that if $\Nself \subseteq
\pc{B} \subseteq \Self$ or $\ORsep \subseteq \pc{B} \subseteq \Nrep$, then
$\FV{B}$ is $\coNP$-complete.

This leaves the $\coNP$-hardness of $\FV{B}$ for $B$ such that $\NRsep
\subseteq \pc{B}$ and the $\DP$-hardness of $\FV{B}$ for all $B$ such
that $\Nsep \subseteq \pc{B}$ to show. We reduce $\FV{\Nrep}$ to
$\FV{\NRsep}$ ($\FV{\BF}$ to $\FV{\Nsep}$, resp.). For that, let $C$
be a circuit over the $\Nrep$-base $\sset{x \wedge (y \vee
\overline{z}), 1}$ (over the $\BF$-base $\sset{x \wedge
\overline{y},1}$, resp.) and let $V$ be the set of variables used in
$C$. Build a
circuit $C'$ by taking a variable $x$ that is not contained in $C$ and
replace every occurrence of $1$ in $C$ by $x$.  Then $C' \wedge x$ is
an $\NRsep$-circuit (an $\Nsep$-circuit, resp.), which can only be
satisfied by assignments that set $x$ to 1. For all these assignments,
$C' \wedge x$ is satisfied if and only if $C$ is satisfied.  Therefore
$(C,V) \in \FV{\Nrep}$ ($(C,V) \in \FV{\BF}$, resp.) if and only if
$(C'\wedge x, V) \in \FV{\NRsep}$ ($(C' \wedge x, V) \in \FV{\Nsep}$,
resp.).
\end{proof}

\medskip 

To obtain a classification for the audit problem, we first 
note the following link between the complexity of the
problem $\EFV{B}$ and the audit problem:

\begin{prop}
\label{prop:efv_audit_relation}
Let $B$ be an arbitrary set of Boolean functions, then
\begin{enumerate}[\em(1)]
 \item $\EFV{B} = \PCSAT{B} \cap \AUDIT{B}$, and
 \item If $\PCSAT{B} \in \P$ then $\EFV{B} \redpeq \AUDIT{B}$.
\end{enumerate}
\end{prop}

The classification now is as follows:

\begin{thm} 
Let $B$ be a finite set of Boolean functions.
\begin{enumerate}[\em(1)]
    \item If $B\subseteq\Mon$ or $B \subseteq \Lin$ or $B \subseteq \Nsep$, then
          $\AUDIT{B}$ is tractable,
    \item else $\AUDIT{B}$ is $\coNP$-complete.
\end{enumerate}
\end{thm}

\proof\hfill
\begin{enumerate}[(1)]
  \item Since $\PCSAT{B} \in \P$ if $\pc{B} \subseteq \Lin$ or $\pc{B}
    \subseteq \Mon$ (Proposition~\ref{prop:Csat}), the claim for
such $B$
    follows from Proposition~\ref{prop:efv_audit_relation} and
    Theorem~\ref{thm:existsFV}.  If $B \subseteq \Nsep$, a
    $B$-circuit is either not satisfiable or has always a frozen
    variable, hence the problem is tractable.
  \item If $B \subseteq \Nrep$ or $B \subseteq \Self$, every
    $B$-circuit is trivially satisfiable. Therefore we can use
    Proposition~\ref{prop:efv_audit_relation} and Theorem~\ref{thm:existsFV}
    again. It remains to show that $\AUDIT{B}$ is $\coNP$-hard if
    $\Nsep \subset \pc{B}$.  We will reduce the $\coNP$-complete problem 
    $\overline{\PCSAT{B}}$ (see Proposition \ref{prop:Csat}) to $\AUDIT{B}$.
The proof runs along the same
    lines as in Theorem~\ref{thm:existsFV}. Take an $B$-circuit
   $C$. Since $\Nsep \subset \pc{B}$ there is a $k$
    with $t_k \in \pc{B}$. Define $C' \define
t_k(C,x_1,\dots,x_k)$,
    where $x_i$ is a variable not occurring in $C$ for $1 \leq i \leq
    k$. If $C$ is not satisfiable then $x_i$ is frozen for $1 \leq i
    \leq k$. If $C$ is satisfiable by an assignment $\alpha$ none of
    the variables from $C$ is frozen in $C'$, since $C'$ can be satisfied by
    setting all the $x_i$'s to 1. Furthermore $x_i$ is not frozen for
    $1 \le i \le k$, because $C'(\alpha 1^{i-1}01^{k-i})=C'(\alpha
    1^{k})=1$.\qed
\end{enumerate}

We finish this section with a classification of the unique satisfiability
problem.

\begin{thm}
Let $B$ be a finite set of Boolean functions.
\begin{enumerate}[\em(1)]
  \item If $\Nsep\subseteq \pc{B}$, then $\USAT{B} \redleq \USAT{\BF}$
  \item else if $\NRsep\subseteq \pc{B} \subseteq \Nrep$, then
    $\USAT{B}$ is $\coNP$-complete
    \item In all other cases $\USAT{B}$ is tractable.
\end{enumerate}
\end{thm}

\proof\hfill
\begin{enumerate}[(1)]
 \item Trivially $\USAT{B} \redl \USAT{\BF}$ for an arbitrary set $B$
  of Boolean functions. According to Proposition \ref{prop:circ-property},
$$ \USAT{\BF}  \redl \USAT{\Nsep\cup\{1\}}.$$
We show that 
$  \USAT{\Nsep\cup\{1\}}  \redl \USAT{\Nsep}$, which in turn will prove that
$\USAT{\BF}\redl \USAT{B}$ for all $B$ such that $\Nsep\subseteq \pc{B}$. Let $C$ be 
a $\Nsep\cup\{1\}$-circuit and let $x$ be a variable not
  occurring in $C$. Let $C''$ be the circuit obtained from $C$ in replacing every 
 occurrence of 1  by $x$. Finally consider   $C' \define C'' \wedge x$. 
 Observe, that since $\wedge\in\Nsep$ the circuit $C'$ is an
  $\Nsep$-circuit and $\#\sat(C) = \#\sat(C')$.
 \item If $B \subseteq \Nrep$ we have $\USAT{B} \redleq
  \overline{\SATstar{B}}$. Hence for all $B\subseteq \Nrep$ holds
  $\USAT{B}\in\coNP$ and therefore the $\coNP$-completeness for all $B$
  with $\NRsep \subseteq \pc{B} \subseteq \Nrep$ follows by Theorem
  \ref{thm:selectSAT}.
 \item For all $B \subseteq \Omsep{2}$ or $B \subseteq \Self$ the
  claim holds because as we have seen before any such circuit has at least
$2^{n-1}$ satisfying assignments. If $B
  \subseteq \Mon$, then an $n$-ary $B$-circuit $C$ has more than one
  satisfying assignment if and only if there is an $i \in
  \sset{1,\dots,n}$ such that $C(1^{i-1}01^{n-i})=1$.  If $B
  \subseteq \Lin$, then the number of satisfying assignments for
  every $B$-circuit can easily be determined using its linear normal
  form.\qed
\end{enumerate}

\section{Enumeration problems}
\label{sect:enumeration}

We now present our results for the enumeration problem.
The analogous problem has been
studied in
the constraint context by Nadia Creignou, Jean-Jacques H\'ebrard,
Henning Schnoor, and Ilka Schnoor in \cite{crhe97,schsch07}.
The counting problem (i.e., determine the number of solutions of a given circuit
has been studied in~\cite{rewa05}).

\begin{thm}\label{theorem:enumeration b formulas:main result}
Let $B$ be a finite set of Boolean
functions. Then the following holds:
\begin{enumerate}[\em(1)]
\item If $B\subseteq\Mon$, or $B\subseteq\Lin$, or $B\subseteq\Self$, or
 $B\subseteq\Omsep 2,$ then $\enumsat{B}$ has a 
  polynomial-delay enumeration   algorithm.
\item Else $\enumsat{B}$  has no  polynomial-total-time enumeration  algorithm unless $\P=\NP$.
\end{enumerate}
\end{thm}

\noindent Note that since every polynomial-delay algorithm is also a polynomial-total-time algorithm, the above theorem
implies that in the context of enumerating the solutions for $B$-formulas, the two notions coincide. In particular,
the theorem completely classifies the ``efficient'' cases with respect to either of these notions.\newpage

\proof\hfill 
\begin{enumerate}[(1)]
 \item Let us first examine the case where $B\subseteq\Mon$ or $B\subseteq\Lin$.
 In this case it follows from Proposition~\ref{prop:Csat} and
 Figure~\ref{platt} that the satisfiability problem for
 $B\cup\set{0,1}$-circuits can be solved in polynomial time. Thus it is easy to see that
the following algorithm has polynomial delay: Let $C(x_1,\dots,x_n)$ be a $B$-circuit. We first check if
 $C[x_1/0]$ (that is, the circuit resulting from $C$ when replacing
 all gates labeled $x_1$ with a gate computing the constant
 $0$-function) is satisfiable, if yes, we recursively print the
 satisfying solutions of this circuit with the additional assignment
 $x_1=0.$ We do the same for the analogously defined $C[x_1/1].$ For a
 circuit without variables, we print the empty assignment.

Let us now consider the case where $B\subseteq\Self$, or
 $B\subseteq\Omsep 2.$ In this case, as we have seen in the proof of Theorem \ref{thm:selectSAT}, we
know that for any $B$-circuit $C$ and any assignment $I$ to the
variables of $C,$ if $I$ is no solution for $C$, then $\dual{I}$
is. This gives a polynomial-delay enumeration algorithm for the
solutions of $C$, by testing the set of all assignments in an
appropriate order: let the variables of $C$ be $x_1,\dots,x_n$, then
use an arbitrary order, for example the lexicographical order, on the
assignments $I$ with $I(x_1)=0$, and for each of the assignments
considered, test if $I$ or $\dual{I}$ satisfies the circuit. In the
cases where the answer is ``yes,'' print the corresponding
assignment. Due to the above mentioned property, this gives at least
one solution for each $I$ considered, since if $I$ is not a solution,
then $\dual{I}$ is. Therefore, since it can be verified in polynomial
time if a given assignment is a solution for the circuit, this clearly
gives a polynomial delay algorithm.
\item   
According to Figure \ref{platt} in order to complete the proof of the theorem 
 it remains to show  that if $B$ is such that $\NRsep\subseteq\clone B$,
then $\enumsat{B}$  has no  polynomial-total-time enumeration  algorithm unless $\P=\NP$.
 We show that the existence of such an algorithm for $B$-circuits
implies that $\SATstar B$ can be decided in polynomial time. The
theorem then follows from the proof of Theorem~\ref{thm:selectSAT},
since there it was proven that $\SATstar B$ is \NP-hard.

Let $C$ be a $B$-circuit. First check if the constant $1$-assignment
is a solution of $C,$ this can be done in polynomial time. Let $i$ be $1$
if this is the case, and let $i$ be $0$ otherwise (i.e., if the constant
$1$-assignment does not satisfy $C$). Clearly,
$C$ has a solution different from the all-$1$-solution if and only if it
has at least $i+1$ many solutions. Using a 
  a polynomial-total-time enumeration algorithm
for $B$-circuits, this question can be decided as follows:

Since $i+1$ can be at most $2$, the time that a polynomial-total-time
enumeration algorithm can spend for enumerating all of $C$'s solution
is bounded by a polynomial in $C$. Therefore, we can simply start the algorithm, and
wait if it finishes in this time. If it does, then its output is the
full list of solutions for $C$, and we obviously can decide if there
is solution different from the constant-$1$-solution present in this
list. If it does not finish in this time, then there are more than $i+1$
solutions, and thus there is one which is not the
constant-$1$-solution. Note that we deduce this fact solely from the observation
that the algorithm runs longer than allowed for $i+1$ solutions, independent of
any output the algorithm may have printed up to that time.\qed
\end{enumerate}

\noindent In the case of the existence of a polynomial-delay enumeration algorithm it 
is of interest to further examine the complexity of the enumeration when requiring 
the solutions to be output in lexicographic order. As observed in \cite{jpy88} this further
requirement can dramatically increase the complexity. We prove that this is indeed the case for
some sets $B$.

\begin{prop}\label{prop:enumlex}
Let $B$ be a finite set of Boolean
functions such that $B\subseteq\Mon$, or $B\subseteq\Lin$, or $B\subseteq\Self$, or
 $B\subseteq\Omsep 2$.
\begin{enumerate}[\em(1)]
\item If $B\subseteq\Mon$  or $B\subseteq\Lin$, then there exists a 
  polynomial-delay enumeration   algorithm that produces all the solutions of a $B$-circuit
in lexicographic order.
\item Else such an algorithm does not exist unless $\P=\NP$.
\end{enumerate}
\end{prop}

\begin{proof}
 Observe that the enumeration algorithm described in the proof of Theorem 
\ref{theorem:enumeration b formulas:main result}
when $B\subseteq\Mon$  or $B\subseteq\Lin$ produces the solutions in lexicographic order.
 According to Figure \ref{platt} it remains to consider the case where
 $\ORsep\subseteq\clone B$ or $\Nself\subseteq\clone B$. We prove that for any set $B$, if one of these algorithms exists, then
the satisfiability problem for $B\cup\set{0}$-circuits can be solved
in polynomial time. The result then follows with 
Proposition~\ref{prop:Csat}, since due to Figure~\ref{platt},
$\clone{B\cup\set{0}}=\BF,$ and therefore this problem is
\NP-complete.

We show how a polynomial-time decision algorithm for this problem can
be obtained from a  polynomial-delay enumeration algorithm for
$B$-circuits that produces solutions in lexicographic order. 
To this end, let $C$ be a $B\cup\set{0}$-circuit. Introduce a new variable $x_0$, and
construct the circuit $C'$, which is obtained from $C$
by replacing every occurrence of 0 by   $x_0$. Then $C'$ is a $B$-circuit. It is clear that
  $C$ has a solution if and only if $C'$ is satisfiable
and the lexicographically first solution of $C'$ maps $x_0$ to $0,$
which clearly finishes the proof, since the lexicographic order
enumeration algorithm has to produce the first solution in polynomial
time, or determine that none exists. 
\end{proof}

 In addition to the cases where the
satisfiability problem for $B$-circuits is \NP-complete, and therefore
efficient enumeration algorithms obviously cannot be hoped for unless
$\P=\NP,$ we also showed that in the cases where tractability of the
satisfiability problem follows from a simple ``trick,'' like the
knowledge that the all-$1$-assignment is a solution to the circuits,
efficient enumeration algorithms do not exist. An interesting special
case here is the case of self-dual circuits. The satisfiability
problem again is easy, simply because any such circuit is always
satisfiable. But the property of self-duality does not only give one
solution, it guarantees that half of the possible assignments are
solutions. Therefore it is not surprising that these solutions also
can be enumerated in an efficient way.  However, since the property of
self-duality does not say anything about the set of solutions where a
given variable is set to $0$, this does not help us to construct a
lexicographical order enumeration algorithm.

Given the above results and those on counting given in~\cite{rewa05},
one can see that counting is ``harder'' than enumeration in the following
sense: For all cases in which~\cite{rewa05} gives a polynomial-time algorithm
for the counting problem, we also obtain an efficient (polynomial-delay) algorithm
for enumeration. The converse is not true: For monotone functions, efficient enumeration
is possible, but counting cannot be done in polynomial time, unless $\#\P\subseteq\FP$.
When considering lexicographic enumeration algorithm, the picture
is similar, with the notable exception of the clone of the self-dual functions:
As already discussed above, enumeration is trivial for these functions. For a similar
reason, the counting problem is trivial here as well (a self-dual function is satisfied
by exactly half of its possible arguments). However, the self-dual property does not
help in obtaining an algorithm for enumeration in lexicographic order.

\section{Conclusion}

\label{sect:conclusion}
We have obtained complete classifications for the equivalence and isomorphism problems,  the frozen variables
problems, the unique satisfiability problem, the audit problem and the
enumeration problem for Boolean circuits. 
The classification into ``hard'' and ``easy'' classes can be refined such that
the internal structure of the tractable cases becomes
visible. For this, one has to use stricter reductions (e.g., logspace reductions
or logtime projections), and one obtains problems complete for subclasses of $\P$.
For some of our problems, this has been done in \cite{rei01}.

We think it is interesting to observe that, e.g., equivalence of OBDDs is decidable in polynomial time, while we identify here intractability for many clones in the lattice. As a consequence, this shows that, if $\P\neq\NP$, in all these cases OBDDs provide a provably less succinct representation than Boolean circuits. Analogous remarks hold for cases of the other algorithmic tasks that we consider, where a difference in complexity between OBDD representation and circuit representation appears.

In general, given a Boolean function $f$, it is $\co\NP$-hard to determine if it is in a clone $B$ (if $f$ is given by a general circuit; the problem becomes very easy if $f$ is given by truth-table, see \cite{Vollmer09}). This might seem to destroy all relevance of our just discussed results. However, we would like to mention that in practice, circuits computing functions $f$ are synthesized in one way or the other, hence we know the minimal clone it belongs to, and thus, our tractability results are relevant.

In this paper we studied the complexity of problems
related to circuits. So it is natural to ask what can be said about
the formula case. For this we define $B$-formulas as ``tree-like''
$B$-circuits or analogously as $B$-circuits, where all gates have a
fan-out of at most $1$. Interestingly the study of $B$-formulas leads
to different dichotomy-theorems. The main reason for this phenomenon
is, that circuits can be regarded as a succinct representation of
formulas. Partial results in this direction have been obtained in
\cite{rei01,DBLP:journals/ijfcs/Schnoor10}.

Finally we would like to mention that another possible syntactic restriction of
formula related problems is to consider generalized Boolean CNF formulas, also
known as CSPs, see \cite{crkhsu00,crvo08}. Many results about the problems
considered here have been obtained in the CSP framework, see the survey
\cite{crvo08}.

\subsubsection*{Acknowledgement.} We are grateful to the reviewers for many  comments that helped to improve the presentation considerably.

\newcommand{\etalchar}[1]{$^{#1}$}

\end{document}